\documentclass{andp2012}
\usepackage{braket}
\usepackage{graphicx}
\usepackage{dcolumn}
\usepackage{bm}
\usepackage{appendix}
\usepackage{subfigure}
\usepackage{hyperref}
\usepackage{color}
\usepackage{xcolor}
\usepackage{amsmath}
\usepackage[english]{babel}

\title{High-speed quantum transducer with a single-photon emitter in a 2D resonator}
\author[F. Author]{Xingyu Gao\inst{1}}
\author[S.\,X. Author]{Zhang-qi Yin\inst{2}}
\author[T.\,Y. Author]{Tongcang Li\inst{1,3,4,5}\footnote{Corresponding author\quad E-mail:~\textsf{tcli@purdue.edu}}}
\address[1]{Department  of  Physics  and  Astronomy, Purdue  University,  West  Lafayette,  IN  47907,  USA}
\address[2]{Center of Quantum Technology Research, School of Physics, Beijing Institute of Technology, Beijing 100081, China}
\address[3]{School  of  Electrical  and  Computer  Engineering, Purdue  University,  West  Lafayette,  IN  47907,  USA}
\address[4]{Purdue Quantum Science and Engineering Institute,  Purdue  University,  West  Lafayette,  IN  47907,  USA}
\address[5]{Birck Nanotechnology Center,  Purdue University,  West Lafayette,  IN 47907,  USA}
\shortauthors{Xingyu Gao et al.}
\begin{abstract}
  Quantum transducers can transfer quantum information between different systems. Microwave-optical photon conversion is important for future quantum networks to interconnect  remote superconducting quantum computers with optical fibers.  Here we propose a high-speed quantum transducer based on a single-photon emitter in an atomically thin membrane resonator that can couple single microwave photons to single optical photons. The 2D resonator is a freestanding van der Waals heterostructure (may consist of  hexagonal boron nitride, graphene, or other 2D materials) that hosts a quantum emitter. The mechanical vibration (phonon) of the 2D resonator interacts with optical photons by shifting the optical transition frequency of the  single-photon emitter with strain or the Stark effect. The mechanical vibration couples to microwave photons by shifting the resonant frequency of a LC circuit that includes the membrane. Thanks to the small mass of the 2D resonator, both the single-photon optomechanical coupling strength and the electromechanical coupling strength can reach strong coupling regime. This provides a way for high-speed quantum state transfer between a  microwave photon, a phonon, and an optical photon. 
\end{abstract}
\shortabstract

\begin{document}
\maketitle

\section{\label{sec:level1}Introduction}
In the past few years, optomechanical and electromechanical systems have gained remarkable attentions for  achieving coherent quantum control \cite{2}. These hybrid devices are leading candidates for transfering quantum information between different forms, such as photonic, phononic, electronic, and spin states \cite{3,4,Bochmann2013,32,33,34,35,36,yin2015b}. In particular, the opto-electro-mechanical coupling of single microwave (or radio-frequency) photons to single optical photons is attractive for future quantum technologies  \cite{Bochmann2013,32,33,34,35,36}. One potential application will be to use optical photons to coherently interconnect remote superconducting quantum computers that use microwave photons. Converting classical microwaves to optical lights has been achieved with metal-coated Si$_3$N$_4$ (or SiN) membrane resonators (thickness $\sim$ 100 nm) \cite{32,33} embedded in LC circuits, and nanobeam piezo-optomechanical crystal cavities (thickness $\sim$ 200 nm) \cite{Bochmann2013,34}.

\begin{figure}[h]
	\centering
	\includegraphics[width=0.46\textwidth]{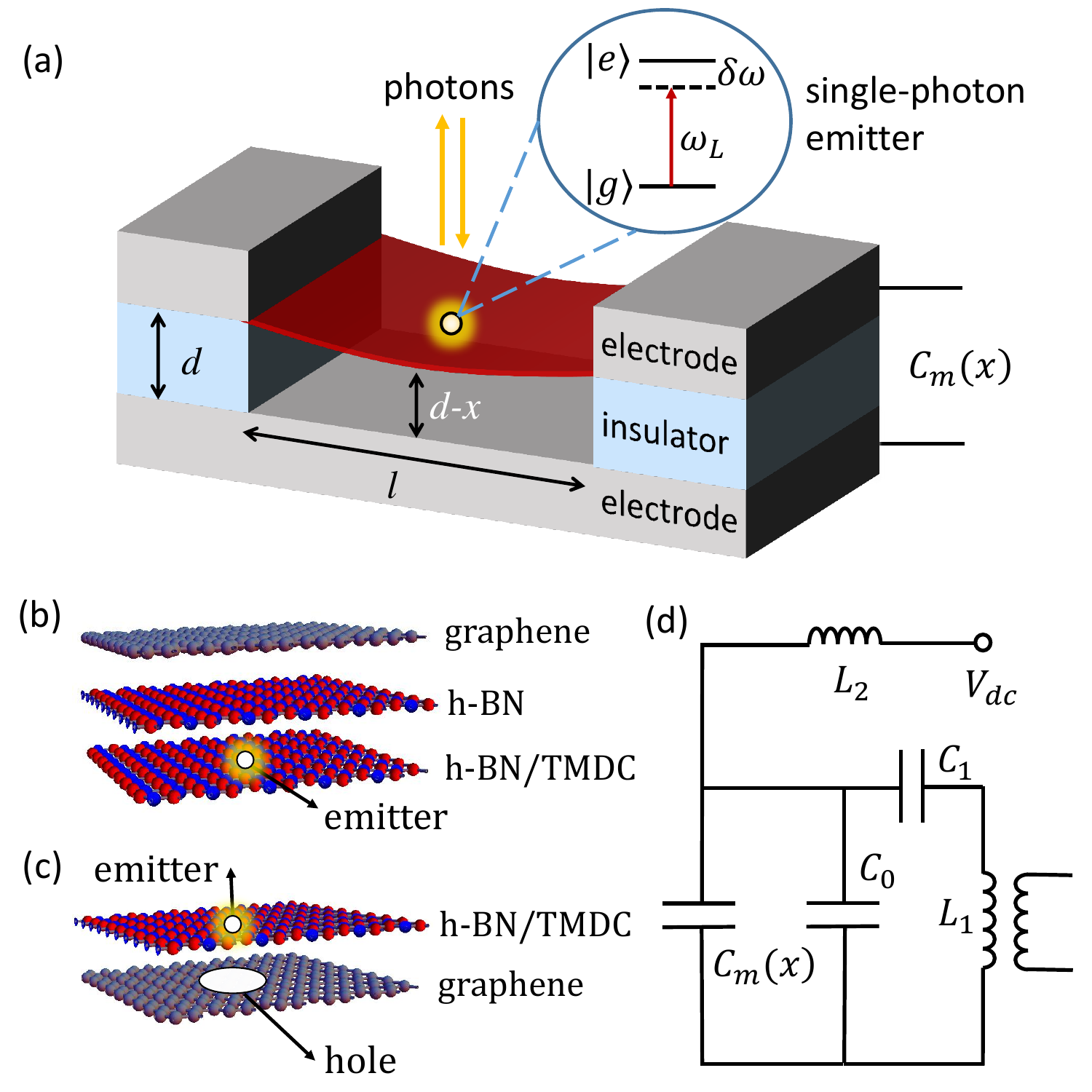}
	\caption{(Color online) Schematic of a high-speed quantum transducer with a single-photon emitter in a 2D resonator. (a) A doubly camped 2D membrane hosting a quantum emitter is suspended to from a mechanical resonator. The device couples to an optical system and an electric microwave resonator by the mechanical vibration of the   2D membrane. $|e\rangle$ and $|g\rangle$ are electron orbital states of the single-photon emitter. The decay from $|e\rangle$ to $|g\rangle$ can emit a visible or infrared photon with frequency $\omega(x)$. (b) An ultrathin 2D membrane contains 3 layers. A top graphene monolayer is connected to the electric circuit. An intermediate h-BN insulating monolayer separates the graphene from the bottom h-BN or TMDC (e.g., WSe$_2$) monolayer which contains a single-photon emitter. (c) A 2D membrane contains 2 layers. The bottom graphene monolayer is connected to the electric circuit . The top h-BN or TMDC monolayer contains a single photon emitter. The graphene has a hole to avoid direct contact of the single-photon emitter with graphene. (d) The 2D membrane capacitor $C_m(x)$ is a part of a LC resonator that contains additional  capacitors $C_{0}$, $C_{1}$ and an inductor $L_{1}$. A constant bias voltage $V_{dc}$ is applied to the membrane to tune the resonant frequency and the coupling strength. A very large inductor $L_{2} \gg L_1$ and a large capacitor $C_{1} \gg  C_{0}+C_m(x))$ are used to isolate the constant bias voltage $V_{dc}$ and the GHz microwave signal in the LC resonator. }\label{fig:1}
\end{figure}

Converting a single microwave photon to a single optical photon with unit fidelity and high speed has been a challenging task. It is difficult to reach strong coupling regime with single photons, which requires the single-photon optomechanical and electromechanical coupling strengths to be larger than the optical decay rate, the mechanical decay rate, and the microwave decay rates \cite{34}. 
In this paper, we propose  a quantum transducer that couples a single microwave photon to a single optical photon with a quantum emitter in a suspended 2D membrane (Fig.\ref{fig:1}). The 2D membrane is a freestanding van der Waals heterostructure. It may consist of graphene, hexagonal boron nitride (h-BN), transition metal dichalcogenide (TMDC), or other 2D materials.
The graphene will be a part of a capacitor in a LC circuit that couples the mechanical vibration (phonon) of the 2D membrane  to microwave photons in the circuit \cite{13,16}.  The h-BN/TMDC layers host a single-photon emitter \cite{27,Vamivakas2015,Imamoglu2015,Potemski2015,26,7}  that couples its mechanical vibration to single optical photons with strain or the Stark effect \cite{14,17,18,57}.   The single-photon emitter replaces the role of the optical cavity in the former opto-electro-mechanical systems \cite{Bochmann2013,32,33,34,35,36}. Here the mechanical vibration of the 2D resonator couples to the electron orbital state of the single-photon emitter, instead of its electron spin state\cite{Rabl2009,31,Yin2013,Ma2016,Ma2017,li2020preparing,xu2019quantum,chen2019nonadiabatic}. Thus it does not require a magnetic field gradient.
We propose to apply a constant voltage to the graphene electrode to increase the strain of the 2D membrane and the charge in the LC circuit to achieve strong coupling regime. The mechanical vibration frequency of the 2D membrane can be tuned by a few GHz with a voltage-controlled strain to match the frequency of a superconducting qubit, which is typically about 5~GHz \cite{kelly2015,zheng2017}.

The proposed quantum transducer with a single-photon emitter in a 2D resonator (Fig.\ref{fig:1}) has several advantages. At low temperatures, the intrinsic linewidth of the zero-phonon line (ZPL) of a single-photon emitter  can be much smaller than the linewidth of a nanoscale optical cavity (typically a few GHz) \cite{34}. At cryogenic temperatures, the nautural linewidth of a quantum emitter in h-BN can be about $30 - 50$ MHz \cite{sontheimer2017photodynamics,dietrich2017narrowband}.

Thanks to the small mass of a 2D resonator,
both  single-photon optomechanical coupling strength and single-photon electromechanical coupling strength of this system can exceed 100 MHz under suitable conditions, which can be larger than the optical decay rate ($\sim$ 40 MHz), mechanical decay rate ($\sim$ 10 kHz), and the microwave decay rate ($\sim$ 10 kHz) of this system \cite{dietrich2017narrowband,19,5,13,will2017high,lee2019ultrahigh}. Thus the system can obtain strong coupling with single photons. 
Ideally, the quantum transducer can be put  together with a superconducting quantum processor in a dilution refrigerator below 50 mK to operate \cite{kelly2015,5,11,12}. Because of the high vibration frequency ($> 1$ GHz), the relevant mechanical mode of a nanoscale 2D resonator will be automatically prepared in the quantum ground state  in a dilution refrigerator.  Thus no optical cooling is required to reach the quantum regime.

Generally, in 2D membranes there would exist a lot of modes which corresponds to different frequencies. To select a typical mechanical mode, we can tune the resonant frequency of the LC circuit to the first bending mode of the membrane and applied a red-detuned laser matching the ZPL of quantum emitters to realize the coupling between the emitter and target bending mode. As other mechanical modes are not coupled to the LC circuit and quantum emitter, they will keep staying at ground state due to the low temperature.

To use this system in a real quantum network, the optical photons from single-photon emitters should be collected efficiently, which is also an active research topic. For example, recently D. Wang, et al used a microcavity with a movable micro-mirror to modify the emission of a single molecule which allows the observation of $99\%$ emitted photons from a single molecule \cite{60}. The micro-mirror of this microcavity was mounted on the tip of a optical fiber, which is applicable in our proposed system. In Ref.\cite{61}, they used a dielectric planar antenna which is a layered structure to change the emission angle of a single molecule which realized a collection efficiency of $96\%$. And there are also some works using fiber to couple single photon source based on solid state emitters \cite{hunger2010fiber,snijders2018fiber}.  All these works can improve the efficiency of photon collection in our proposed system making it more practical in quantum network. 

In the following parts of this paper, we will first describe the model in section \ref{sec:level10}, and discuss the mechanical vibration feature in section \ref{sec:level2}. The electromechanical coupling and optomechanical coupling based on the strain and the Stark effect will be estimated in section \ref{sec:level12}. In section \ref{sec:level5}, we simulate a scheme of high speed quantum state transfer from a microwave photon to an optical photon, which demonstrates that a high fidelity can be achieved through optical readout. In the last section, we briefly summarize the results of the paper.

\section{Model} \label{sec:level10}
\subsection{Scheme} \label{subsec:Model_scheme}
As show in Fig.\ref{fig:1}, we consider a quantum transducer with a freestanding van der Waals heterostructure consisting of graphene and hexagonal boron nitride (and/or a TMDC monolayer) that hosts a quantum emitter. The electron orbital state of a single photon emitter embedded in this membrane couples to the mechanical bending mode $\hat{b}$. Therefore, the spontaneous emission spectrum (around $600$ nm) of the single photon emitter is dependent on the mechanical mode $\hat{b}$.
The mechanical oscillator also forms a capacitor with the bottom electrode, which couples the mechanical mode $\hat{b}$ to the electric microwave mode $\hat{c}$.

Our proposal is based on recent  progresses in atomically thin  memchanical resonators (e.g., a suspended graphene membrane)  \cite{5}, and single-photon emitters in TMDC monolayers \cite{27,Vamivakas2015,Imamoglu2015,Potemski2015} and h-BN \cite{26,7,ahn2018stable}.
Suspended 2D membranes have outstanding mechanical properties such as ultrahigh Young's modulus and the ability to sustain remarkable strain without breaking \cite{6,15}. These mechanical systems can be precisely engineered with high quality factors \cite{11,38,12,39}. Their extremely small mass and out-of-plane stiffness give them a large zero-point vibration amplitude for strong coupling.  h-BN  has a bandgap of about 6 eV, and can host single photon emitters deep inside its bandgap, which is stable even at room temperature \cite{26,7}. These quantum emitters exhibit narrow
ZPL distributed over a large range from 550 nm to 800 nm \cite{dietrich2017narrowband}. Monolayer tungsten diselenide (WSe$_{2}$) has a direct bandgap of about 1.9 eV, and can host localized single photon emitters at low temperature \cite{27,Vamivakas2015,Imamoglu2015,Potemski2015}.
Optical properties and energy-levels of single-photon emitters in 2D materials can be modulated by applying strains or external fields \cite{7,8,31}. For example, bright and photostable single-photon emitters embedded in hexagonal boron nitride  shows great spectra tunability under strain control \cite{7}. Localized emitters in a monolayer WSe$_{2}$  are demonstrated to have efficient spectra tunability by the Stark effect \cite{8}. It is remarkable that a 5-layer van der Waals heterostructure consisting graphene, h-BN, WSe$_{2}$, h-BN, and graphene has been assembled to measure the Stark effect \cite{8}.  These emitters are stable during material transfer. Recently, emitters with giant Stark effect in h-BN have also been observed at room temperature \cite{xia2019room}. The spectral shift up to 43 meV/(V/nm) was achieved in that work. These properties make it possible to realize ultra strong optomechanical and electromechanical coupling with single-photon emitters in a suspended 2D resonator.   Our proposal only requires a 3-layer or 2-layer freestanding van der Waals heterostructure (Fig.\ref{fig:1}). Similar heterostructure 2D resonators have been demonstrated at room temperature already \cite{ye2017}. Their quality factor will be much higher at low temperatures.

An example of the 2D membrane is shown in Fig.\ref{fig:1}(b). In this example,  a top graphene monolayer is connected to the electric circuit and forms a capacitor with the bottom electrode (Fig.\ref{fig:1}(a)) . An intermediate h-BN insulating monolayer separates the graphene from the bottom h-BN or TMDC (e.g., WSe$_2$, MoS$_2$) monolayer which contains a single-photon emitter.
At low temperature ($T <$ 50mK), the quality factor of the mechanical resonator is very high \cite{5,6}. The electronic excited state $\ket{e}$ of the single-photon emitter in this 2D membrane couples strongly to the lattice strain and the external electric field \cite{7,8,14}.  The lattice strain will change when the membrane vibrates. If we apply a bias voltage $V_{dc}$ to the graphene electrode, the electric field will be $V_{dc}/(d-x)$, where $d$ is the thickness of the insulator, and $x$ is the displacement of the membrane (Fig.\ref{fig:1}(a)). $x$ is negative when the membrane is attracted to the bottom electrode.  The vibration of the membrane will change the electric field.
Thus the single photon emitter can be coupled to the mechanical bending mode $\hat{b}$ by the lattice strain or the Stark effect.
Finally, the single photon from the spontaneous emission of the electron spin couples with the mechanical mode $\hat{b}$.

Another possible membrane containing 2 atomic layers is shown in Fig.\ref{fig:1}(c). The top layer hosts a single photon emitter to couple the optical photon and mechanical phonon through strain effect. The bottom layer is used to apply electric force onto the membrane to tune its strain and resonant frequency. A hole is fabricated on the bottom graphene monolayer to minimize the absorption of photons by graphene \cite{25}.

A driving laser at the red sideband of the electron spin's two-level optical transition is applied to induce resonant coupling between the electron and the mechanical mode $\hat{b}$, and read out the quantum state optically. The frequency of the driving laser is $\omega_L=\omega_0-\omega_m$, where $\omega_0$ is the optical resonant frequency of the single photon  emitter and $\omega_m$ is the vibration frequency of the 2D membrane.

The suspended membrane also forms a capacitor which couples the mechanical vibration to  microwave photons at the frequency $\omega_{LC}$.  As shown in Fig.\ref{fig:1}(d), the 2D membrane capacitor $C_m(x)$ is a part of a LC resonator that contains additional  capacitors C$_{0}$, C$_{1}$ and an inductor L$_{1}$.
Since $C_m(x)$ depends on the position of the 2D membrane,  the mechanical mode $\hat{b}$  couples to the microwave mode $\hat{c}$.
A constant bias voltage $V_{dc}$ is applied to the membrane to tune the coupling strength. A very large inductor $L_{2} \gg  L_1$   and a large capacitor $C_{1} \gg  (C_{0}+C_m(x))$ are used to isolate the constant bias voltage $V_{dc}$ and the high-frequency microwave signal in the LC resonator.  The large inductor $L_{2}$ prevents the GHz microwave photon from leaking to the $V_{dc}$ connector. The capacitor $C_1$ prevents a DC current  passing the $L_1$ inductor.
The resonant frequency of the LC circuit is $\omega_{LC} \approx 1/\sqrt{L_1 (C_m(x)+C_0)}$. We  assume that the coupling rates are much less than the mode spacing of these resonators so that only  one mode is relevant for each degree of freedom.

\subsection{Hamiltonian} \label{subsec:Model_Hamiltonian}
The Hamiltonian of the system takes the form \cite{14,13,16}
\begin{align}
\label{H}H=\hbar \omega(x) \ket{e}\bra{e}+ \frac{p ^2}{2m}+\frac{m\omega_m^2 x^2}{2}+\frac{\phi ^2}{2L}+\frac{q^2}{2C(x)}-qV_{dc} 
\end{align}
\noindent where $\phi$ and $q$  correspond to the flux in the inductor and the charge on the capacitors, respectively. For the  excited state $\ket{e}$ of the single photon emitter, the electron-phonon interaction leads to the energy shift of the zero photon line (ZPL), which results in the dependence of the frequency $\omega(x)$ of the ZPL on the displacement of the membrane $x$.  Here we introduce the phonon creation(annihilation) operator $b^{\dagger}$($b$) and microwave photon creation(annihilation) operator $c^{\dagger}$($c$)  , with
\begin{eqnarray}\label{eq:opterators}
\begin{split}
&x=x_{zpf}(b+b^{\dagger}),\;p=im_{eff}\omega_{m}x_{zpf}(b-b^\dagger)\\
&q=q_{zpf}(c+c^{\dagger}),\;\phi=iL\omega_{LC}q_{zpf}(c-c^\dagger)
\end{split}
\end{eqnarray}
\noindent where $x_{zpf}=\sqrt{\hbar/(2m_{eff}\omega_{m})}$, $q_{zpf}=\sqrt{\hbar/(2L\omega_{LC})}$ are zero-point fluctuations of the mechanical mode and the microwave mode, respectively. Then the Hamiltonian can be is given by $H=H_{0}+H_{I}$, where (see details in Appendix A)
\begin{eqnarray}
&\label{H0}H_{0}=\hbar \omega_{0} \ket{e}\bra{e} +\hbar \omega_{m} b^\dagger b+\hbar \omega_{LC} c^\dagger c,\\
&H_{I}=\hbar g_{om}(b+b^{\dagger}) \ket{e}\bra{e}+g_{em}(b+b^{\dagger})(c+c^{\dagger}).
\end{eqnarray}
\noindent Here $\omega_0$ is the frequency of the electron spin when $x$ equals to $\bar{x}$ and $\bar{x}$ is the equilibrium position of the membrane.  $g_{om}$ and $g_{em}$ are the optomechanical and electromechancial coupling rates, respectively.  The eletromechanical coupling strength takes the form $g_{em}=G x_{zpf} q_{zpf}$ and $G=\bar q\frac{\partial }{\partial x}(\frac{1}{C(x)})|_{x=\bar x}$. And the optomechincal coupling strength is given by $g_{om}=x_{zpf}(\partial \omega_{0}/\partial x)$ (See Appendix A). More details about the optomechanical and electromechancial coupling strengths will be discussed later in Section \ref*{sec:level12}.

In the real experiment, we need to consider decays in the system. There are three decay channels: The quantum emitters embedded in 2D membranes couple to the environment through optical decay channel; The mechanical resonator and superconducting circuit also introduce noise due to finite quality factors Q. Here we assume that the optical, mechanical and electrical damping rates are $\kappa$ ($\sim 2 \pi \cdot$40 MHz), $\Gamma_{m}$ ($\sim 2 \pi \cdot$10 kHz) and $\Gamma_{LC}$ ($\sim$ $2 \pi \cdot$ 10 kHz) respectively \cite{19,5,13}. To achieve strong coupling, $g_om$ and $g_em$ should be larger than $\kappa$, $\Gamma_{m}$ and $\Gamma_{LC}$. While strong cooperativity (defined in the next subsection) is usually thought to be enough for a quantum transducer, our simulation will show that achieving strong coupling regime will further improve the quality of quantum  transducer (Section \ref{Fidelity}), which allows near unit fidelity and high converting speed. Thus, in the following calculation of coupling strength, we will first give the cooperativity of each coupling term and then show feasibility of achieving strong coupling regime.

\begin{figure}[bp]
	\centering
	\includegraphics[width=0.49\textwidth]{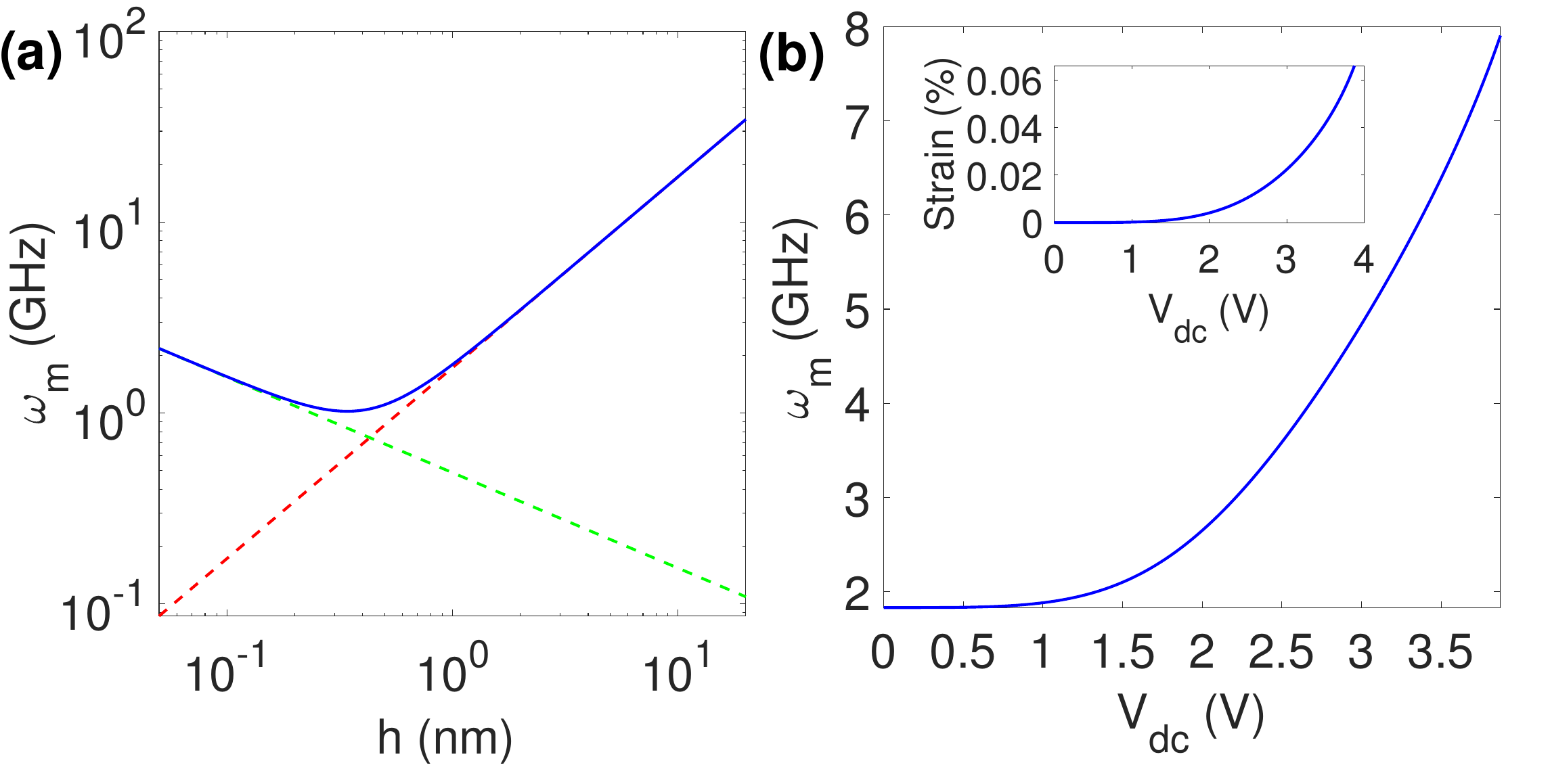}
	\caption{(Color online) (a) Thickness dependence of the resonant frequency of a 2D membrane without a bias voltage (solid blue line). The effective Young's modulus is set as 867 GPa and the pre-tension is $T=10$ nN \cite{5,9}. The length and width of the membrane are assumed to be $l=110$nm and $w=1 \mu$m, respectively.  The thickness of a real 2D membrane is discrete, corresponds to certain points in this plot. A transition from flexible membrane (yellow dashed line) to stiff plate  (red dashed line) mechanical behaviors can be observed in this figure.   (b) Voltage dependence of the resonance frequency of a 2D membrane. The thickness is taken as 1 nm, corresponding to 3 layers. A bias voltage leads to a displacement of the membrane which increases the tension $T$.  Other parameters are the same as those in (a).}\label{fig:6}
\end{figure}

\section{ Mechanical Vibration}\label{sec:level2}

The mechanical resonator (Fig. \ref{fig:1}) that we discuss here is based on  doubly clamped  ultrathin 2D materials like graphene, h-BN or WSe$_{2}$ \cite{23,24}. These materials have ultrahigh Young's modulus ($\sim1000$ GPa for graphene and 800 GPa for h-BN) and can sustain strains up to 25\% without breaking.
Experiments have examined the relation between the frequency of the fundamental mechanical bending mode and the dimensions of the 2D membrane \cite{5,9}. For mechanical resonators under tension $T$, the fundamental flexural mode frequency $\omega_{m}$ is given by \cite{5,9}:
\begin{equation}
\label{frequency} \frac{\omega_{m}}{2\pi}=\left( \frac{A^2 Y h^2}{\rho l^4}+\frac{0.57 A^{2}T}{\rho l^{2}wh} \right)^{1/2},
\end{equation}
\noindent where $w$, $l$ and $h$ are the width, length and thickness of the structure, respectively. $A$ is the clamping coefficient, which is 1.03 for doubly clamped membranes \cite{6}. $Y$ is the Young's modulus and $T$ is the  pre-tension of the suspended 2D materials. The typical initial tension of a doubly clamped suspended graphene has been measured in Ref.\cite{6} to be about 0.01 N/m. The frequency of the fundamental flexural mode as a function of the  thickness of the 2D membrane is shown in Fig.\ref{fig:6}(a). The length and width of the membrane are assumed to be $l=110$nm and $w=1 \mu$m, respectively. The thickness of a real 2D membrane is discrete, corresponds to certain points in this plot. A transition from flexible membrane (yellow dashed line) to stiff plate  (red dashed line) mechanical behaviors can be observed in this figure. For very thin layers, the frequency in Eq.(\ref{frequency}) is dominated by the second term which is determined by pre-tension $T$. For a thick flake, the tensile strain induced by vibration is more important than the pre-tension. So the first term in Eq.(\ref{frequency}) will dominate.

In the following discussions, the typical dimensions of the mechanical membrane are assumed to be ($l$, $w$, $h$)=(110~nm, 1~$\mu$m, 1~nm) \cite{5,falin2017mechanical}. The frequency of the fundamental flexural mode when there is no bias voltage  is calculated to be about 1.83 GHz, and the zero-point fluctuation amplitude is about 0.14~pm. The initial distance between the membrane and the capacitor chip is important in order to achieve high coupling rate at small DC bias voltages. Here we assume the initial distance to be $d=10$~nm. A bias voltage $V_{dc}$ can be tuned to achieve optimum coupling rates.

The elastic properties of freestanding 2D materials have been studied in several experiments. The relationship between the force and the deformation $\delta$ at the center of the doubly clamped structure is \cite{5,29}
\begin{equation}
\label{C1}F=\left[\frac{30.78wh^{3}}{l^{3}}Y+\frac{12.32}{l}T \right]\delta+\frac{8whY}{3L^{3}}\delta^{3}.
\end{equation}
Because the membrane is very thin, a bias voltage will induce an electric field that can cause a dramatic deflection of the membrane. For a doubly clamped suspended membrane with a parallel bottom electrode that form a capacitor, the electrostatic force as a function of the bias voltage is given by \cite{5,30}
\begin{equation}
\label{C2}	F=\frac{\epsilon_{0} w l V^{2}_{dc}}{2(d-x)^2}.
\end{equation}
The equilibrium position can be derived from Eq.(\ref{C1}) and Eq.(\ref{C2}).
The deformation will increase the tension $T$  and change the mechanical resonance frequency as showed in Fig\ref{fig:6}(b). In this system, we can tune the mechanical resonance frequency by adjusting the bias voltage to match the frequency of a superconducting qubit. Typically, the frequency of a superconducting circuit is around 5 GHz \cite{37,39}. A  3.2 V bias voltage will cause a $x=2.4$ nm displacement and can shift the mechanical vibration frequency from 1.83 GHz to 5 GHz to match the frequency of a superconducting qubit.

\begin{figure}[h]
	\centering
	\includegraphics[width=0.5\textwidth]{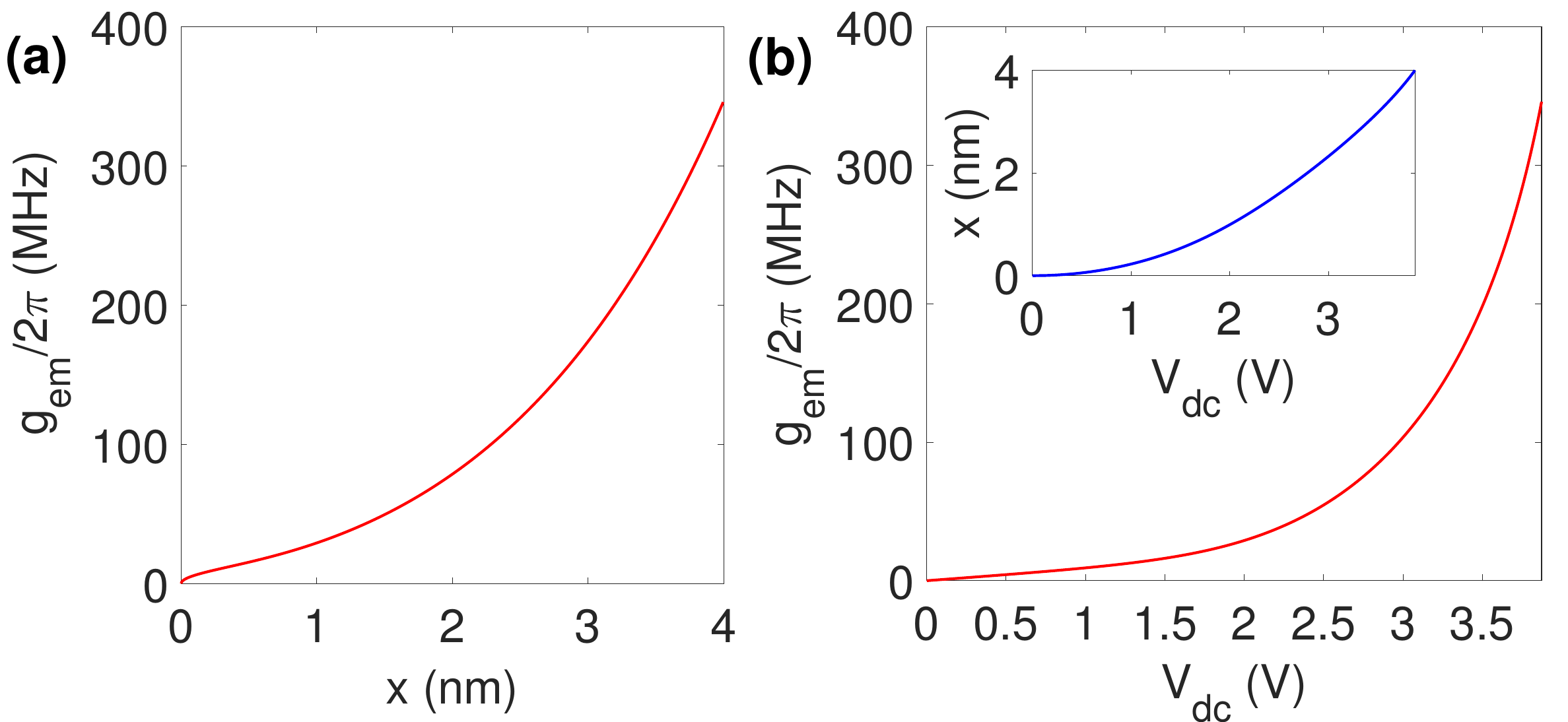}
	\caption{(Color online) The coupling strength ($g_{em}/2\pi$) between a microwave photon in the LC circuit and  the mechanical vibration of the 2D membrane as a function of the displacement of the 2D membrane (a)  and (b) the bias voltage.  Here the displacement is caused by the bias voltage, as shown in the inset of subfigure (b). We assume the parameters of the 2D membrane are $l=110$ nm, $w=1 \mu$m and $h=1$ nm. The initial distance between the center of membrane and the bottom electrode is $d=10$ nm. The  inductance is taken as L=1 $\mu$H. The capacitance $C_0$ is tuned to match the resonant frequency of the 2D membrane.}\label{fig:2}
\end{figure}

\section{Coupling Strength} \label{sec:level12}
\subsection{Electromechanical Coupling}\label{sec:level3}
As shown in Fig. \ref{fig:1}, the 2D membrane and the bottom electrode forms a capacitor $C_{{m}}(x)$.  Its capacitance depends on the separation ($d-x$) between the 2D membrane and the bottom electrode. The vibration of the 2D membrane changes the value of $C_{{m}}(x)$, and thus couples to the microwave photon in the LC circuit.  With another paralleled tuning capacitor $C_{0}$, the total capacitance is $C(x) \cong C_{0}+C_{m}(x)$. Here we assume $C_1 \gg C_{0}+C_{m}(x)$. So the effect of $C_1$ on the high frequency microwave signal can be neglected.

As discussed in the former section, a bias voltage $V_{dc}$ will be applied to the capacitor to tune the mechanical vibration frequency $\omega_{m}$ of  the 2D membrane to match the LC circuit's resonance frequency $\omega_{LC}$.
The bias voltage will charge the capacitors  $\bar q=V_{dc}C(x)$. In this case, the electromechanical coupling between the  microwave photon and the mechanical phonon can be described by the Hamiltonian
\begin{equation}
H_{em}=\hbar g_{em}(b+b^{\dagger})(c+c^{\dagger})
\end{equation}
\noindent where $g_{em}=G x_{zpf} q_{zpf}$ and $G=\bar q\frac{\partial }{\partial x}(\frac{1}{C(x)})|_{x=\bar x}$. Thus the electromechanical coupling strength depends on the parameter G, which is proportional to the bias voltage $V_{dc}$ and   $C'_{m}(x)/C(x)$.

The electromechanical coupling works when the LC circuit's resonant frequency $\omega_{LC}$  matches the mechanical vibration frequency  $\omega_{m}$. Assuming the inductor is $L_1=1$ $\mu$H \cite{37,39,40} and the quality factor is $Q_{LC}=50000$, the total capacitor should be $C=1.3$ fF to have a resonant frequency of 5 GHz. The electric damping rate will $\Gamma_{LC}=\omega_{LC}/Q_{LC} = 2 \pi \cdot 10$ kHz.

The relationship between the electromechanical coupling rate ($g_{em}/2\pi$) and the  bias voltage is displayed in Fig.\ref{fig:2}. When the bias voltage increases, the predisplacement $x$ of the membrane and the charge  $\bar q$ will increase correspondingly. Thus the coupling rate improves. In this case, we find the single microwave photon-phonon coupling rate can be about 150 MHz when the bias voltage is 3.3 V.  We assume $l=110$ nm, the width $w=1 \mu$m and the thickness $h=1$ nm (corresponds to 3 layers). The graphene-based ultrathin mechanical resonator can achieve very high quality factor ($>$10000) \cite{39,5} in low temperature ($<$50 mK) \cite{5,11}. If the mechanical damping rate is 100kHz, the electromechanical cooperativity will be $c_{em} \equiv \frac{g^{2}_{em}}{\Gamma_{LC}\Gamma_{m}}=3.9\times 10^{6}$, which is far larger than 1.

\begin{figure}[h]
	\centering
	\includegraphics[width=0.5\textwidth]{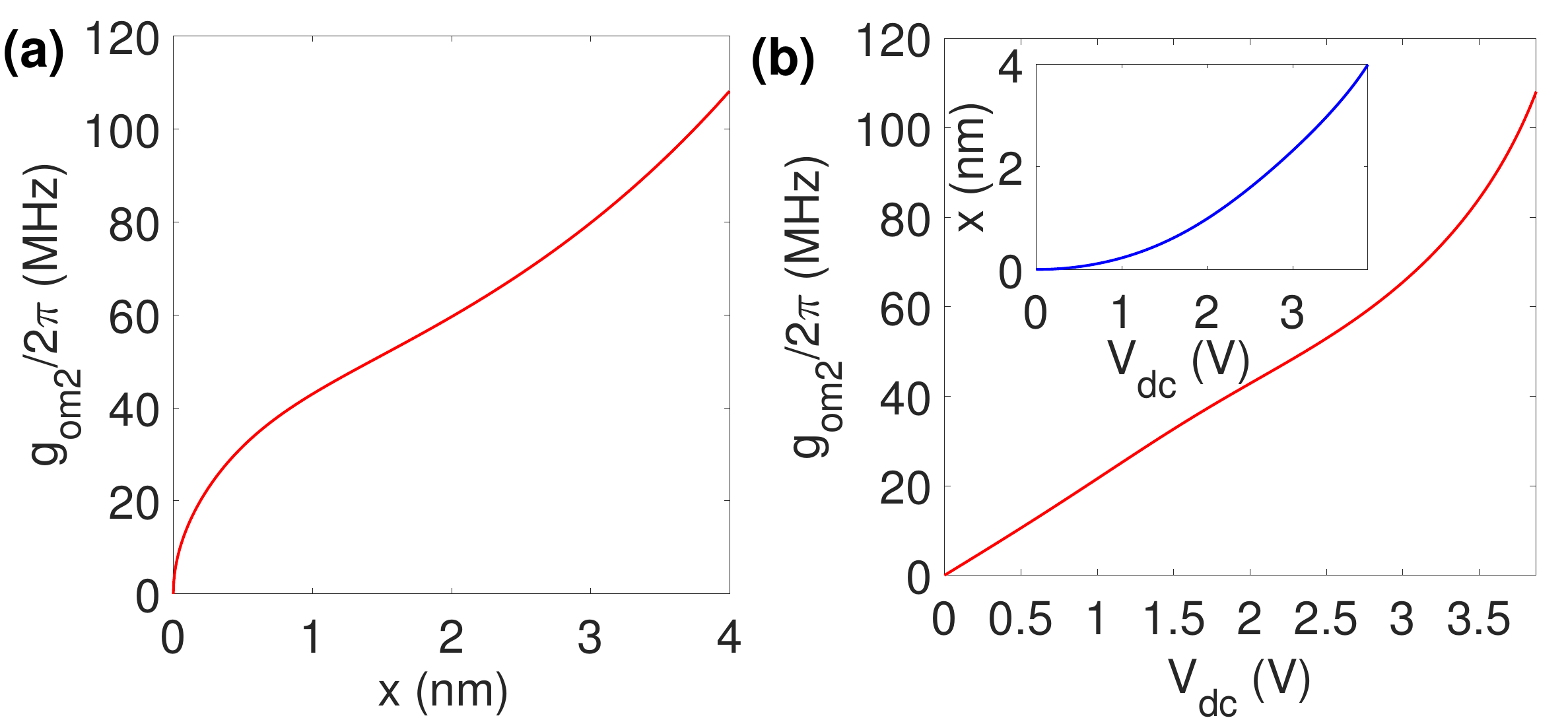}
	\caption{(Color online) The Stark effcet-induced optomechanical coupling strength ($g_{om2}/2\pi$) between an  electron orbital states of the single-photon emitter and the mechanical vibration of a 2D membrane as a function of (a) the displacement  and (b) the bias voltage. The displacement is caused by the bias voltage as shown in the inset of (b). Parameters of the 2D membrane are the same as in Fig. \ref{fig:2}. }\label{fig:5}
\end{figure}

\subsection{Optomechanical Coupling}\label{sec:level4}
In cavity optomechanics, the mechanical vibration of a mirror couples to photons by changing the length and the resonant frequency of a cavity \cite{2}. In our case, the mechanical vibration of a membrane will change the lattice constant of the membrane, and thus shifts the frequency $\omega(x)$ of the optical transition of a single photon emitter in the 2D membrane.  So the mechanical phonon of the 2D membrane can be coupled to  optical photons by the strain effect. The zero phonon line (ZPL) of a single photon emitter in a 2D membrane is also  sensitive to electric fields due to the Stark effect.  In our proposed device (Fig. \ref{fig:1}), the vibration of the 2D membrane will change the electric field $V_{dc}/(d-x)$ between the membrane and the bottom electrode, and thus can couple to optical transitions of the single photon emitter  by the Stark effect.
The coupling  between a mechanical vibration mode of the 2D membrane and the quantum emitter's electronic state can be described by a Hamiltonian
\begin{align}
H_{om}=\hbar g_{om}(b+b^\dagger)\ket{e}\bra{e}.
\end{align}
Here $g_{om}$=$x_{zpf}\partial \omega_0(x) / \partial x$ is the optomechanical coupling strength. We call it optomechanical coupling because a red sideband  photon at frequency $\omega_L=\omega_0-\omega_m$ is involved in this process, although it is not explicit in this Hamiltonian. The electronic transition between $\ket{g}$ and $\ket{e}$ will happen by absorbing a phonon at frequency  $\omega_m$ and a photon at $\omega_L$ together.

\subsubsection{The Stark Effect-Induced Optomechanical Coupling}

With a bias voltage $V_{dc}$, there will be an electric field at the location of the single photon emitter. The Stark effect of localized emitters in 2D materials has been studied recently. For an emitter in h-BN, a sepctral shift up to 43 meV/(V/nm) was observed \cite{xia2019room}. In our proposed system, the ZPF of the 2D membrane induces an electric field variation $\Delta E_{zpf}=x_{zpf}\partial E/\partial x=x_{zpf}V_{dc}/d^{2}$, which can be quite high ($\sim$10 kV/m). The coupling rate between the optical transition and the mechanical vibration induced by the Stark effect is given by
\begin{equation}
g_{om2}=x_{zpf}\frac{\partial \omega}{\partial x}=\Delta E_{zpf}\frac{\partial \omega}{\partial E}.
\end{equation}
The Stark effect-induced coupling rate $g_{om2}$ is much larger than the strain induced coupling rate $g_{om1}$. The Stark effect-induced coupling rate $g_{om2}$ is plotted as a function of the displacement and  bias voltage in Fig.\ref{fig:5}. The optomechanical coupling rate is larger than 50 MHz when the bias voltage exceeds 2.5 V, and thus the system can obtain ultra strong optomechanical cooperativity. Assuming the decay rate the the excited state $\ket{e}$ is 32 MHz and the decay rate of the mechanical oscillator is 10 kHz, the optomechanical cooperativity is
$c_{om2}=g^{2}_{om2}/(\Gamma_{m}\kappa) \sim 7.8\times 10^3$ when the bias voltage is 2.5V.

\begin{figure}[h]
	\centering
	\includegraphics[width=0.49\textwidth]{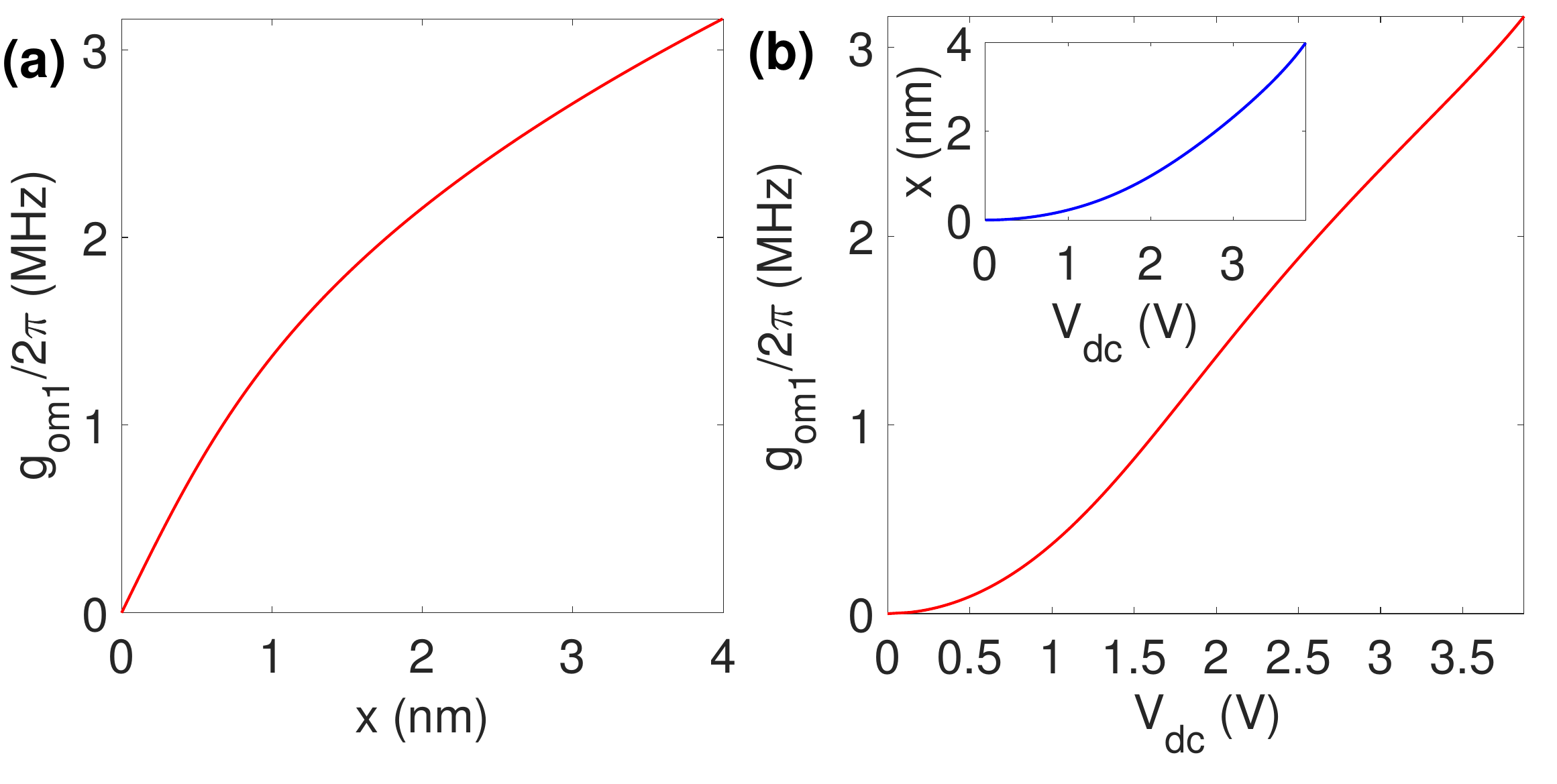}
	\caption{(Color online) The strain-induced optomechanical coupling strength ($g_{om1}/2\pi$) between an  electron orbital states of the single-photon emitter and the mechanical vibration of a 2D membrane as a function of (a) the displacement  and (b) the bias voltage. Parameters are the same as in Fig. \ref{fig:2}. As will be discussed later, the strain-mediated coupling strength shown in this figure is small compared to the Stark effect induced coupling in Fig. \ref{fig:5}. }\label{fig:3}
\end{figure}

\subsubsection{Strain-Mediated Optomechanical Coupling}
In this part, we consider a strain along the length direction caused by the bending of the doubly clamped membrane which has $A_{1}$ symmetry and preserves the $C_{3\upsilon}$ sysmetry of a defect center. An A$_{1}$-symmetric strain preserves the degeneracy of E$_{x}$ and E$_{y}$ orbital states and uniformly shifts the energy of both states. Thus under a A$_{1}$-symmetric strain, the ZPL will have a frequency shift without splitting. Experiments have investigated the spectral shift of single photon emitters in h-BN under different applied strain fields \cite{7}. The studied emitters exhibited a wide span of emission energy shifts from -3 meV per 1$\%$ strain (meV/$\%$) to 6 meV/$\%$ due to preexisting local strains. To  evaluate the optomechanical coupling between the ZPL of a quantum emitter and the mechanical motion of a 2D membrane, we assume the shift coefficient of the ZPL under a strain field to be 5 meV/$\%$.

The strain $S$ applied on the 2D membrane is related to the relative elongation of the length of the membrane: $S=\frac{\Delta l}{l}$, where $\Delta l$ is the length elongation and $l$ is the initial length of the membrane.
If there is no initial deformation of the membrane,  the strain fluctuation induced by the zero point fluctuation of the membrane is negligible because the elongation $\Delta l=2(\sqrt{(l/2)^2+x^2_{zpf}}-l/2) \approx 2 x^2_{zpf}/l $ is a small  second-order function of the the zero point fluctuation  $x_{zpf}$. With a bias voltage, there will be an initial deformation $x_0$ of the membrane which can be calculated from the equations (\ref{C1}) and (\ref{C2}). Then the strain effect $S=\frac{\Delta l}{l}\approx 4x_{0} x_{zpf} /l^2$, which is  2$x_{0}/x_{zpf}$ times of $2 x^2_{zpf}/l $.  Here $x_{0} \sim$ 2 nm can be more than one thousand times larger than $x_{zpf} = 0.14$pm. In this case, the strain mediated coupling strength is
\begin{equation}
g_{om1}=x_{zpf}\frac{\partial \omega}{\partial x}=\frac{4x_{0}x_{zpf}}{l^2}\frac{\partial \omega}{\partial S},
\end{equation}
where $S$ is the strain induced by the mechanical vibration.  Here we calculate the strain-mediated coupling strength based on a single emitter in h-BN. As shown in Fig.\ref{fig:3},  the single excitation coupling rate exceeds $2\pi \times 3$ MHz for a 2D membrane ($l=110$ nm, $w=1 \mu$ m and $h=1$ nm) when its initial displacement is 4 nm. This is relatively large compared to existing experiments \cite{Bochmann2013, 34}. The optomechanical cooperativity due to the strain effect is $c_{om1}=g^{2}_{om1}/(\Gamma_{m}\kappa) \sim 1.8$.

However, $g_{om1}$ is still  smaller than the typical decay rate of the excited state of a single-photon emitter in a 2D membrane.  The strain effect is not large enough to obtain strong coupling. On the other hand, as has been discussed in the previous section, the Stark effect  is large enough to reach the strong coupling regime.

\section{\label{sec:level5} High Speed Quantum State Transfer} \label{Fidelity}
Most of optomechanical devices are hitherto in the limit of the weak single-photon optomechanical coupling \\ regime, $g_{om} \ll \kappa$ \cite{Bochmann2013,32,33,34,35,36}. In that case, a strong driving field is required to enhance the effective coupling strength, which results in a large average photon number and a low signal-to-noise ratio. Our proposed system uses a single-photon emitter in a 2D membrane replacing the optical cavity to  mediate the coupling between an optical photon and a mechanical phonon. Thanks to the small mass of 2D membranes and the relatively narrow ZPL of a single-photon emitter at low temperature, it is much easier to obtain ultra strong cooperativity. The single-photon coupling rate can easily exceed 100 MHz, making it possible to realize high speed quantum state transfer under a relatively weak laser driving field.

\subsection{Scheme}

Here we consider the quantum  state transfer from a microwave photon in the LC circuit to an optical single-photon pulse output. We assume that the system works at low temperature ($<$50 mK). So the electron orbital state $\ket{\psi_{e}}$ of the single photon emitter and the mechanical vibrational mode $\ket{\psi_{m}}$ of the 2D membrane are initialized at ground states, i.e., $\ket{\psi_{e}(t=0)}$=$\ket{0}$ and $\ket{\psi_{m}(t=0)}$=$\ket{0}$. Then the LC resonator is prepared into state $\ket{\psi_{LC}(t=0)}$=$\ket{1}$. The process aims to transfer the quantum state from $\ket{\psi_{e}(t=0)} \otimes \ket{\psi_{m}(t=0)} \otimes  \ket{\psi_{LC}(t=0)}$ =$\ket{001}$ to $\ket{100}$. As spontaneous emission couples electron orbital state to the free-space continuum optical modes, once the electron state evolves into excited state $\ket{e}$, an optical photon may be spontaneously emitted from the transition $\ket{e}$ to $\ket{g}$ and the system will output a single-photon pulse. To simplify the calculation, as the coupling strengths in the system are much stronger than the mechanical and electric damping rates, we first derive the effective Hamiltonian by ignoring the decay and then take it back into consideration during the second part. Here a laser driving field is required. Its  effect  can be described by a Hamiltonian
\begin{equation}
H_{d}=\hbar\frac{\Omega}{2} (e^{-i\omega_{L}t}\ket{e}\bra{g}+e^{i\omega_{L}t}\ket{g}\bra{e}),
\end{equation}
where $\omega_{L}$ is the frequency of the driving laser and $\Omega$ is the Rabi frequency. Then, applying the Schrieffer-Wolff transformation
$U=\exp[\frac{g_{om}}{\omega_{m}}(b^\dagger-b)\ket{e}\bra{e}]$ to the total Hamiltonian and keeping only the near resonant terms, we obtain the effective Hamiltonian in the interaction picture (see details in Appendix B)
\begin{equation}
\tilde{H_{I}}=\hbar \tilde{g}_{om}(b \sigma_{eg}+b^\dagger \sigma_{ge})+\hbar g_{em}(bc^\dagger+b^\dagger c),
\end{equation}
where $\tilde{g}_{om}=\frac{\Omega}{2}\frac{g_{om}}{\omega_{m}}$ is the effective coupling strength between the electronic state of the single photon emitter and the mechanical vibration mode. We assume $\tilde{g}_{om}=g_{em}=g$ for simplicity of notation. The wave function takes the form $\ket{\psi}=\ket{\psi_{e}}\ket{\psi_{m}}\ket{\psi_{LC}}$. If the system is initialized into the state  $\ket{\psi_{e}}\ket{\psi_{m}}\ket{\psi_{LC}}=\ket{g}\ket{01}$, the quantum state will evolute within the subspace $\ket{g}\ket{01}$, $\ket{g}\ket{10}$ and $\ket{e}\ket{00}$. This Hamiltonian has three eigenstates which are $\ket{\psi_{1}}=\frac{1}{2}(\ket{g}\ket{01}-\sqrt{2}\ket{g}\ket{10}+\ket{e}\ket{00})$, $\ket{\psi_{2}}=\frac{1}{2}(\ket{g}\ket{01}+\sqrt{2}\ket{g}\ket{10}+\ket{e}\ket{00})$ and $\ket{\psi_{3}}=\frac{\sqrt{2}}{2}(\ket{g}\ket{01}$\\$-\ket{e}\ket{00})$.
Then we get the quantum state of this system
\begin{align}
\ket{\psi(t)}=&\frac{1}{2}e^{i\sqrt{2}gt}\ket{\psi_{1}}+\frac{1}{2}e^{-i\sqrt{2}gt}\ket{\psi_{2}}+\frac{\sqrt{2}}{2}\ket{\psi_{3}}\\
\begin{split}
=&\frac{1}{2}(1+\cos{\sqrt{2}gt})\ket{g}\ket{01}\\
&-\frac{1}{2}(1-\cos{\sqrt{2}gt})\ket{e}\ket{00}\\
&-i\frac{\sqrt{2}}{2}\sin{\sqrt{2}gt}\ket{g}\ket{10}
\end{split}	
\end{align}

In the above simplified discussion, we neglect the coupling of the electronic transition $\ket{g}{\leftrightarrow }\ket{e}$ to the optical photon output and other decay channels. Now we consider three decay channels that couples to this system: the optical channel with decay rate $\kappa$, the mechanical channel with damping rate $\Gamma_{m}$ and the electric channel with damping rate $\Gamma_{LC}$. At non-zero temperature, the decoherence in this system is accelerated due to the stimulated emission, and the decay rates are proportional to the number of thermal quanta. Thus, the effective decay rate is given by $ (n_{i}+1)\kappa$ \cite{36} and the number of thermal quanta in each channal is given by $n_{i}=1/(e^{\hbar\omega_{i}/k_{B}T}-1)$. Here $T$ is the temperature of the environment.  At the temperature of $50$ mK, $\bar{n}_{0}\approx 0$, $\bar{n}_{m}= 0.008$ and $\bar{n}_{LC}= 0.008$ with $\omega_0=\omega_{LC}=2\pi \cdot 5$GHz.   The natural linewidth of emitters in h-BN in this case can be as narrow as 32 MHz \cite{sontheimer2017photodynamics}.

The photon output from the system is around the ZPL $\omega_0$ of single photon emitters, which has a frequency difference $\omega_{m}$ from the frequency of the driving laser $\omega_L$. This photon output is the result of the microwave-optical photon conversion. It means that the optical decay channel contains the information of the quantum state transfer, rather than leading to the loss in the process. The main channels leading to quantum decoherence are the mechanical dissipation and electric circuit dissipation, which are quite small since $\Gamma_m$ and $\Gamma_{LC}$ are much smaller than $\kappa$.

To qualify the coupling between the transition $\ket{g}{\leftrightarrow }\ket{e}$ and single photon output through optical decay channel,  we introduce $a$ to denote the one-dimensional free-space photon modes which couples to the atomic transition with coupling strength $\sqrt{\kappa/2\pi}$. Using the method in Ref. \cite{10}, the whole conditional Hamiltonian \cite{RevModPhys.70.101,PhysRevA.94.022302}, including the mechanical decay and electric circuit decay, can be written in the following form in the rotating frame:
\begin{align}
\begin{split}
H_{I}=&\hbar g(b \ket{e}\bra{g}+b^\dagger \ket{g}\bra{e})+\hbar g(bc^\dagger+b^\dagger c)\\
&+i\hbar\sqrt{\kappa/2\pi}\int_{-\omega_{a}}^{\omega_{a}	 }d\omega(\ket{e}\bra{g}a(\omega)-\ket{g}\bra{e}a^\dagger(\omega)\\ &+\int_{-\omega_{a}}^{\omega_{a}}d\omega[\hbar\omega a^\dagger(\omega)a(\omega) ]-i\hbar\frac{\Gamma_{m}}{2}b^\dagger b-i\hbar\frac{\Gamma_{LC}}{2}c^\dagger c. \label{dd}
\end{split}
\end{align}
To solve the dynamics governed by the Hamilton (\ref{dd}), we can expand the state $\ket{\psi}$ of the whole system into the following superposition:
\begin{align}
\begin{split}
\ket{\psi}=&(c_1 \ket{e}\ket{00}+c_2 \ket{g}\ket{10}+c_3 \ket{g}\ket{01})\otimes\ket{vac}\\
&+\ket{g}\ket{00}\otimes\ket{\phi}, \label{qs}
\end{split}
\end{align}
where $\ket{vac}$ denotes the vacuum state of the free-space photon mode $a$, and
\begin{align}
\ket{\phi}=\int_{-\omega_{a}}^{\omega_{a}}c_{\omega}a^\dagger(\omega)\ket{vac} d\omega
\end{align}
represents the state (not normalized) of the single-photon output pulse. The coefficients $c_1$, $c_2$, $c_3$ and $c_\omega$  are time dependent. At  time $t=0$, we have $c_1=0$, $c_2=0$, $c_3=1$ and $c_{\omega_{j}}=0$.

After applying a red sideband driving laser at $\omega_L$, these coefficients changes with $t$. We need to compute the time evolution of these coefficients by substituting $\ket{\psi}$ into the Schr{\"{o}}dinger equation $i\hbar\partial_{t}\ket{\psi}=H\ket{\psi}$. For numerical representation of the Hamiltonian (\ref{dd}), we discretize the free-space field $a(\omega)$ by introducing a finite but small frequency interval $\delta \omega$ between two adjacent modes. Then,  we have about  $N=2\omega_{a}/\delta\omega$ free-space modes in total. The $j$-th mode is denoted by $a_{j}$ whose frequency detuning from the central frequency is given by $\omega_{j}=(j-N/2)\delta\omega$. Here $\delta\omega$ is much smaller than the inverse of the total evolution time $T$. The integral bandwidth $\omega_{a}$ is much larger than the optical decay rate $\kappa$, but is much smaller than $\omega_{m}$ to guarantee that there will be no change of the physical result by discretization. We rewrite the single-photon state as
\begin{align}
\ket{\phi}=\sum_{j=1}^{N}c_{\omega_{j}}a_{j}^{\dagger}\ket{vac}.
\end{align}
Then we can obtain the following  set of equations for coefficients $c_1$, $c_2$, $c_3$ and $c_{\omega_{j}}$:
\begin{align}
\begin{split}
\dot{c_1}=&-igc_2 +\kappa '\sum_{j=1}^{N} c_{\omega_{j}},
\end{split} \label{c1}\\
\begin{split}
\dot{c_2}=&-igc_1 -igc_3 -\frac{\Gamma_{m}}{2}c_2,
\end{split}	\label{c2}\\
\dot{c_3}=&-igc_2 -\frac{\Gamma_{LC}}{2}c_3,\\
\dot{c_{\omega_{j}}}=&-\kappa 'c_1-i(j-N/2)\delta\omega c_{\omega_{j}},\label{cj}
\end{align}
where the effective optical decay rate $\kappa'\equiv\sqrt{\kappa\delta\omega/2\pi}$. We numerically integrating Eqs. (\ref{c1})-(\ref{cj}) to obtain the solutions.

\begin{figure}[t]
	\centering
	\includegraphics[width=0.48\textwidth]{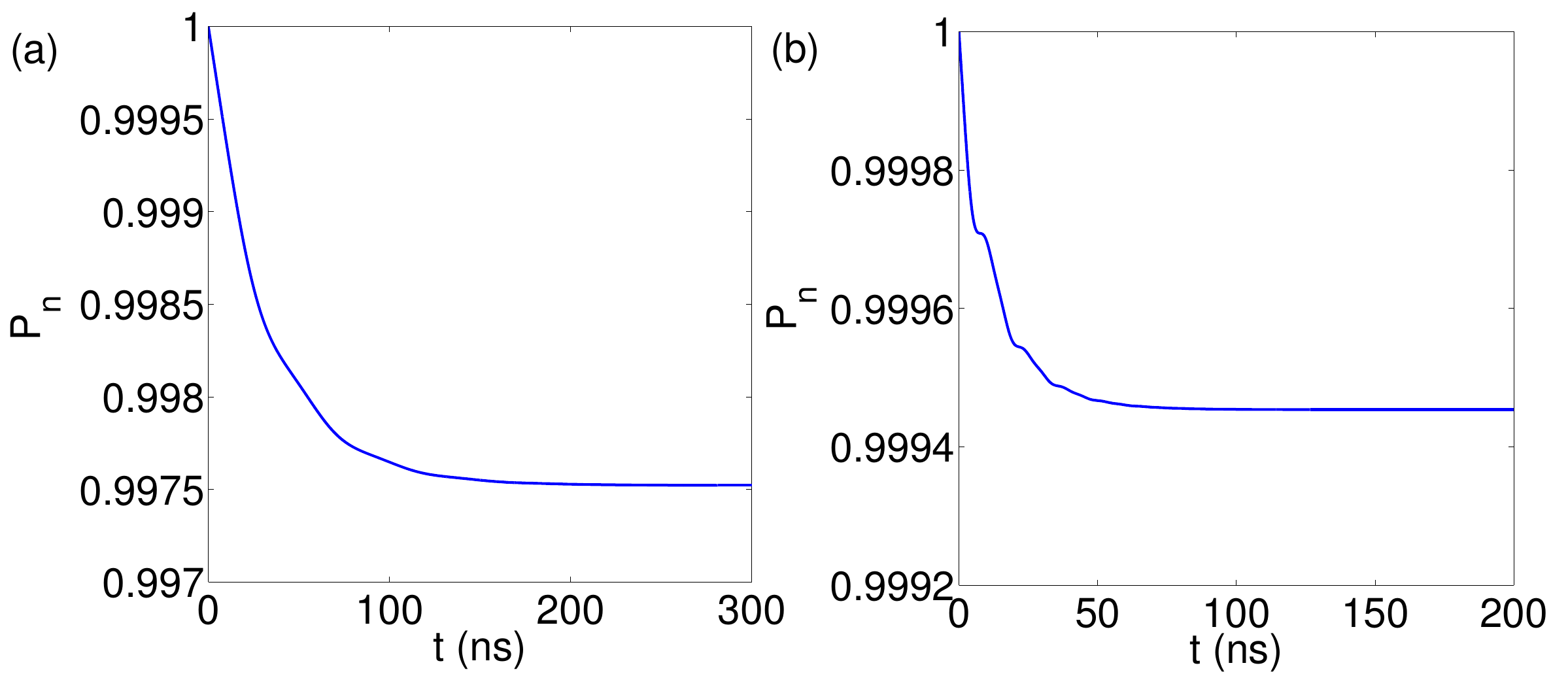}
	\caption{(Color online) The probability $P_n$ of state transfer with no mechanical or electronic loss at time t. The coupling rates are (a) $g/2\pi=10$ MHz and (b) $g/2\pi=50$ MHz, respectively. Other  parameters are  $\kappa/2\pi$=40 MHz, $\Gamma_{m}/2\pi$=10 kHz, $\Gamma_{LC}/2\pi$=10 kHz, and temperature $T=50$ mK. In the numerical simulations, the intervals and numbers of the free-space modes are taken as $\delta \omega/2\pi=0.5$ MHz, N=1000. The results in the figure are independent of the exact values of $\delta \omega$ and $N$ when  $\delta \omega$ is sufficiently small and $N$ is sufficiently large.}\label{fig:c1}
\end{figure}

\subsection{Fidelity} \label{sec:Fidelity}

In this system, the property we concern most is the fidelity of converting a microwave photon in the LC circuit to an optical single-photon pulse output. If the total quantum state is a pure state, the whole process of the quantum state transfer from electric circuit to an optical photon output is reversible, which means that the information in the electric circuit can be totally transferred to optical photons.  However, the Hamiltonian (\ref{dd}) is not Hermitian because of the mechanical and electrical circuit decay terms $-i\hbar\frac{\Gamma_{m}}{2} b^\dagger b$ and $-i\hbar\frac{\Gamma_{LC}}{2} c^\dagger c$. Because of these two decay terms, the initial signal may be lost due to  thermal dissipation by the 2D resonator or the LC circuit, which may lead to no optical photon output. To quantify  the influence of these leakages, we  introduce the probability of quantum state transfer without leakage through the mechanical or LC decay channel:
\begin{align} \label{Pn}
P_{n}=|c_{1}(t)|^2+|c_{2}(t)|^2+|c_{3}(t)|^2+\sum_{j=1}^{N}|c_{\omega_{j}(t)}|^2,
\end{align}
which is the normalization coefficient of the total quantum state at time $t$. Fig.\ref{fig:c1} displays the calculated $P_t$ for two different coupling strengths  $g/2\pi=10$ MHz and  $g/2\pi=50$ MHz. The decay rates of each channel are $\kappa/2\pi$=40 MHz, $\Gamma_{m}/2\pi$=10 kHz, $\Gamma_{LC}/2\pi$=10 kHz. And the temperature is assumed to be $T=50$ mK. The probability $P_t$ for quantum state transfered without mechanical or electric circuit loss is larger than 99$\%$ for both cases.

\begin{figure}[t]
	\centering
	\begin{minipage}{0.5\textwidth}
		\centering
		\includegraphics[scale=0.25]{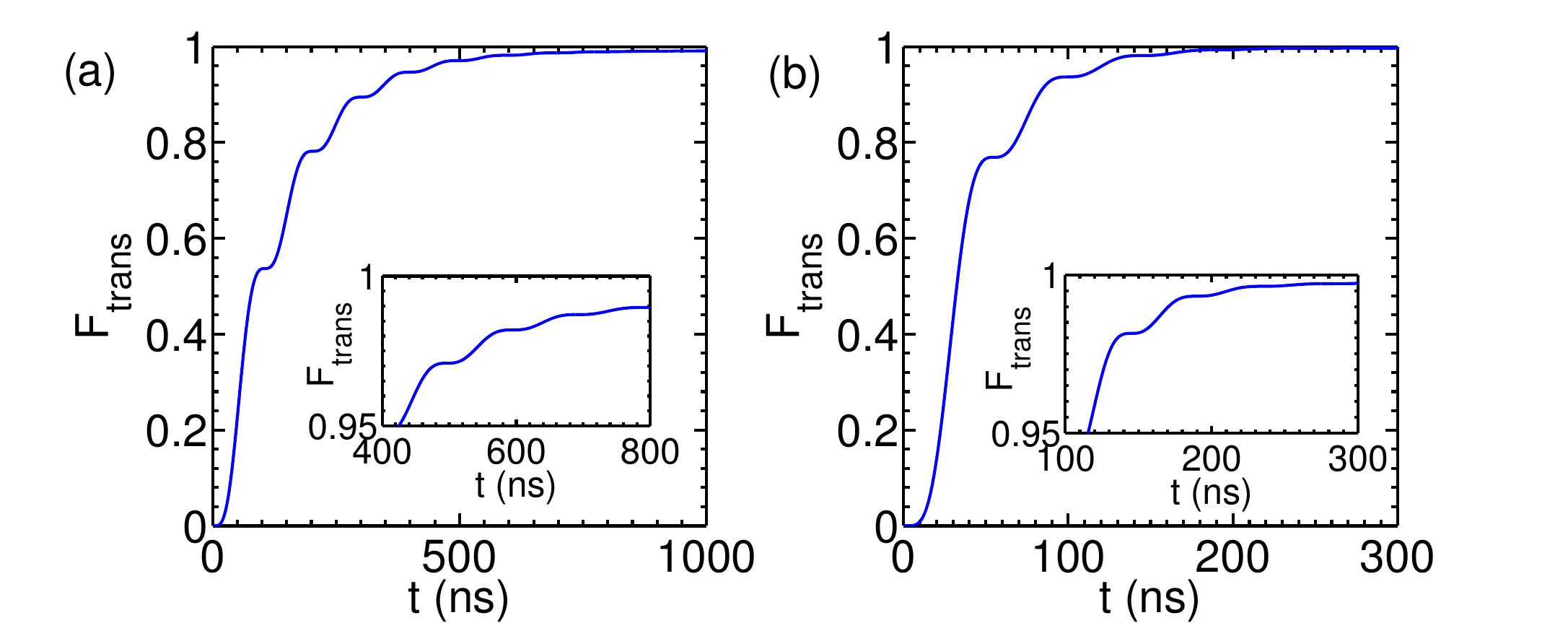}
	\end{minipage}
	
	\begin{minipage}{0.5\textwidth}
		\centering
		\includegraphics[scale=0.25]{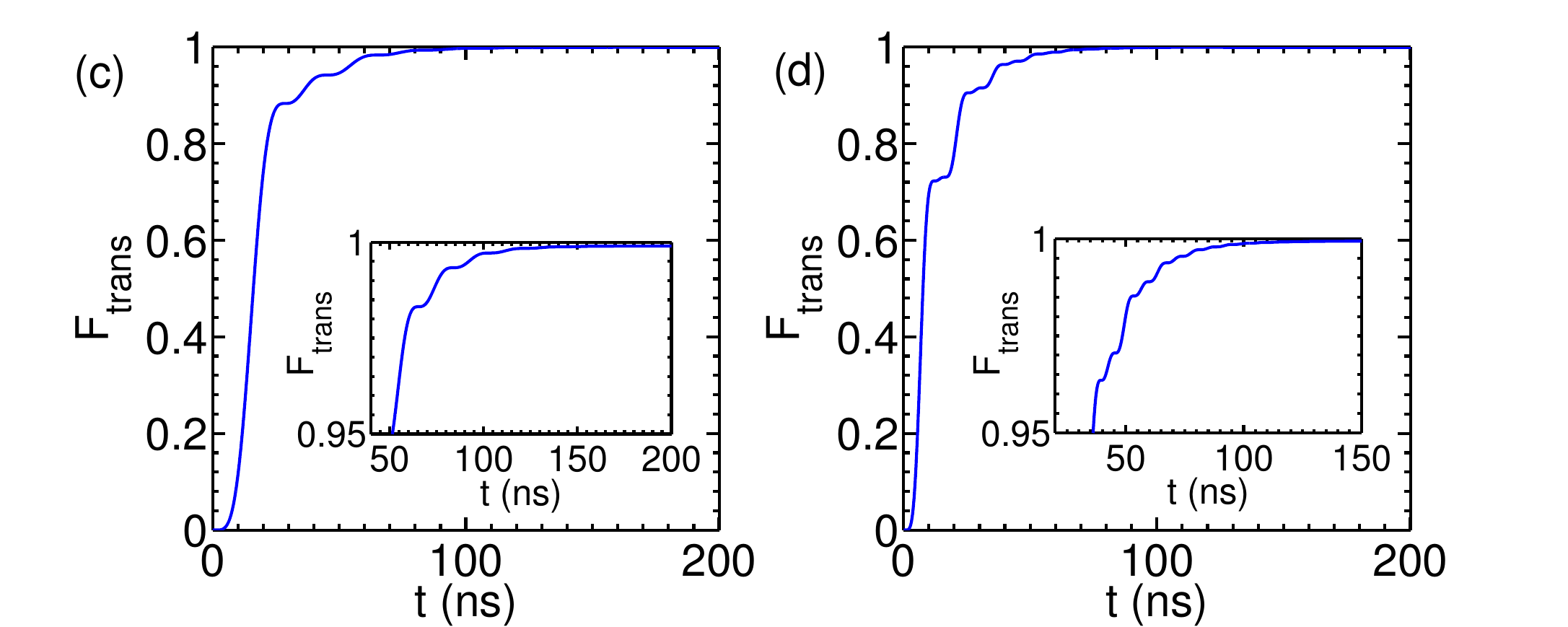}
	\end{minipage}
	\caption{(Color online) Time evolution of the fidelity $F_{trans}$ of the  state transfer from a microwave photon in the LC circuit to a free-space optical photon output for different coupling rates (a) $g/2\pi$=5 MHz, (b) $g/2\pi$=10 MHz, (c) $g/2\pi$=20 MHz, (d) $g/2\pi$=50 MHz. The maximum fidelities are (a) $F_{trans}$=0.9911, (b) $F_{trans}$=0.9973, (c) $F_{trans}$=0.9990 (d) $F_{trans}$=0.9995. Other parameters are taken as temperature $T$=50 mK,  $\kappa/2\pi$=40 MHz, $\Gamma_{m}/2\pi$=10 kHz, $\Gamma_{LC}/2\pi$=10 kHz.
	}\label{fig:c3}
\end{figure}

\begin{figure}[h]
	\centering
	\includegraphics[width=0.48\textwidth]{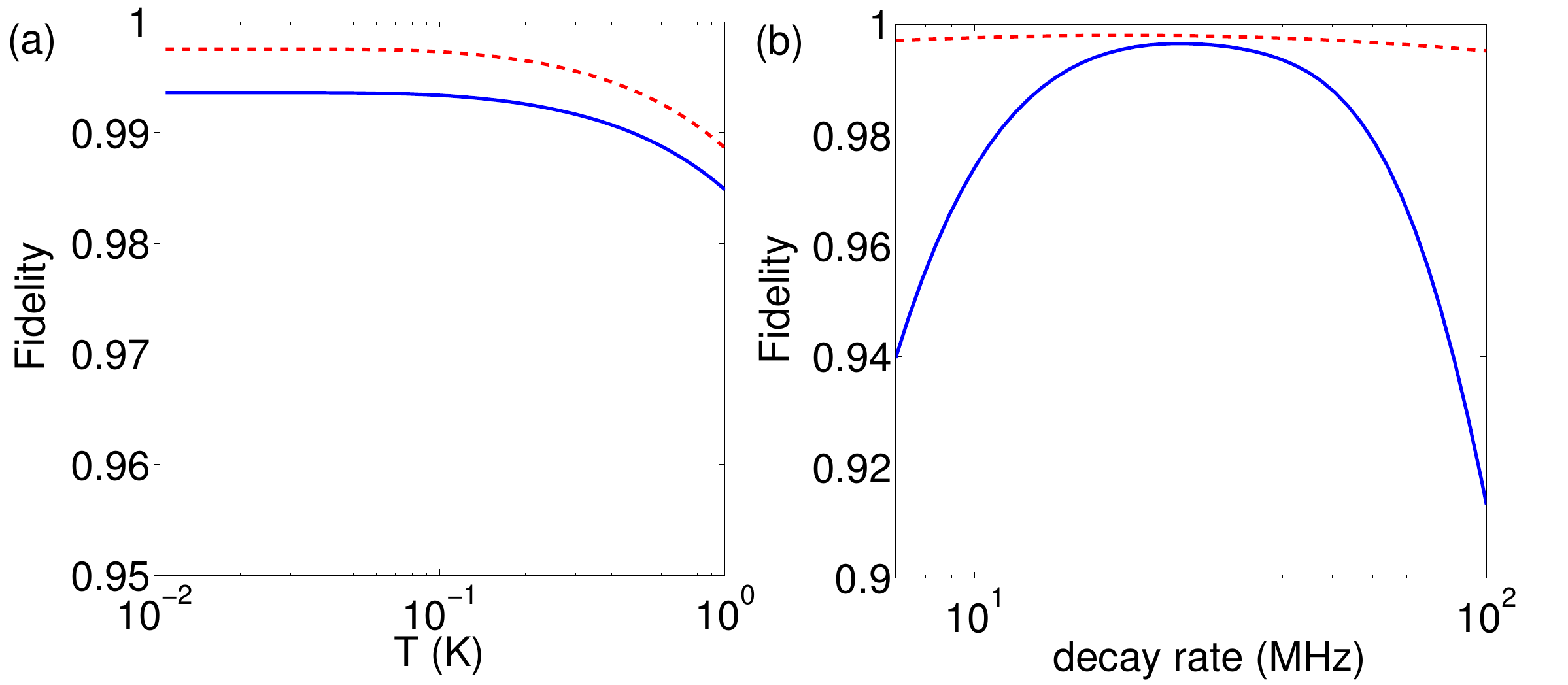}
	\caption{(Color online) (a) Temperature dependence of the fidelity $F_{trans}$ (blue solid curve) of transferring a microwave photon to an optical photon pulse output, and the probability $P_n$ of state transferring with no loss through mechanical and LC channels (red dashed curve). The evolution time is taken as 200 ns. Other parameters are $g/2\pi$=10 MHz, $\kappa/2\pi$=40 MHz, $\Gamma_{m}/2\pi$=10 kHz, $\Gamma_{LC}/2\pi$=10 kHz.   (b)  Dependence of the fidelity $F_{trans}$ (blue solid curve) and the probability $P_n$ (red dash line) on the atomic decay rate $\kappa$ at 50 mK. Other parameters are the same as in (a). } \label{fig:c4}
\end{figure}

The single-photon  output through the optical decay channel is our signal, while all other dissipation channels contribute to the loss.  Here we quantify the state-transfer fidelity  by $F_{trans}$, which is the normalization coefficient of the single-photon pulse state at the time $t$
\begin{align}
F_{trans}=\sum_{j=1}^{N}|c_{\omega_{j}}(t)|^2.
\end{align}
In Fig.\ref{fig:c3} we show the fidelity as a function of the evolution time $t$ with different coupling rates. The fidelity increases with evolving time, and eventually reaches its maximum when the fidelity $F_{trans}$ equals to the probability $P_n$. It is found that when the effective coupling rate greatly exceeds other decay rates,   $G\sim\kappa\gg\Gamma_{m},\Gamma_{LC}$, the fidelity exceeds 99$\%$ as shown in Fig.\ref{fig:c3}(a)-(d). The dependence of  $P_n$ and  $F_{trans}$ on the system temperature and the decay rate $\kappa$ is provided in Fig.\ref{fig:c4}. It shows that the fidelity can maintain a high value within a large scale of the temperature. However, to maintain high fidelity, the decay rate should be close to the couple strength. Here we have ignored the effect from dark counts which will be further discussed in the next section. 

\subsection{Dark Counts}\label{darkcount}

Above we showed that the system can convert a single microwave photon to an optical photon output with high fidelity in a short time. However, there can also be some imperfections in this scheme causing errors in photon counts which may affect the quality of transduction. This means that the transducer may also output photons through optical channel in the absence of a signal microwave photon (dark counts). To analyze the quantity of dark counts, we mainly focus on two possible sources.

\begin{figure}[h]
	\centering
	\begin{minipage}{0.5\textwidth}
		\centering
		\includegraphics[scale=0.21]{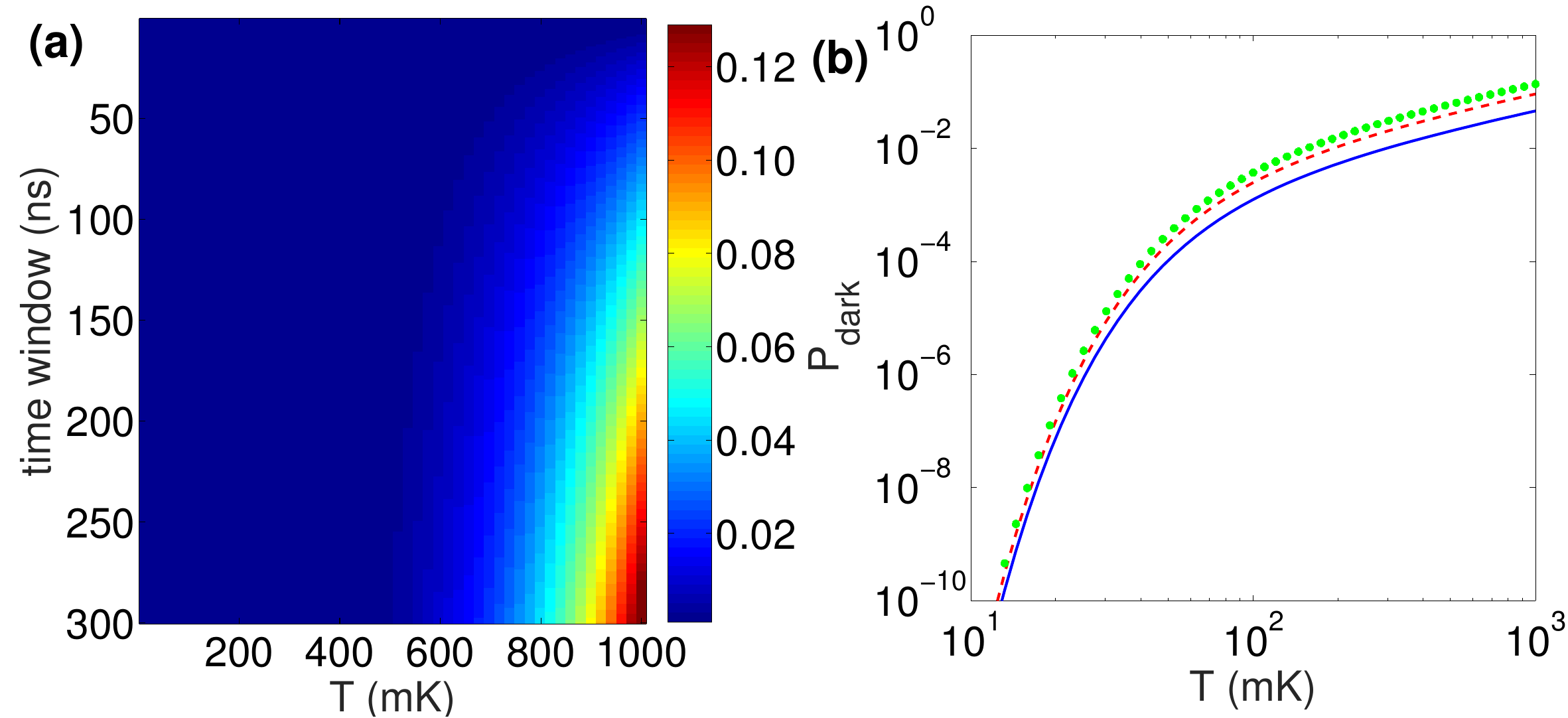}
	\end{minipage}

	\caption{(Color online) Here we present the temperature T and time window dependence of the dark count $n_{th}$ due to thermal excitation. (a) shows the dark count ratios within the readout time window. (b) gives temperature dependence of the dark counts ratios in log scale, which shows that the dark counts from thermal excitation is extremely low when the temperature is lower than 100 mK. The probabilities of a dark count $P_{dark}$ are plotted at three different readout time windows: 300 ns (green dot curve), 200 ns (red dash curve) and 100 ns (blue solid curve).}\label{fig:darkth}
\end{figure}

First, in above discussion, we ignored the thermal excitation due to the non-zero temperature. When the system is coupled to a high-temperature bath, it will be heated continuously through mechanical and electric dissipation channels. Then if a microwave photon (or mechanical phonon) is excited by the surrounding environment, it will be mostly converted into an optical photon due to the near unit transduction fidelity in our system. It means that all those unexpected thermal phonons (photons) during the readout windows will contribute to the bad counts. As the dissipation procedure highly depends on the surrounding temperature, we calculate the temperature dependence of the dark count $n_{th}$. Fig.\ref{fig:darkth} shows that decreasing temperature can greatly depress the dark counts at a fixed readout window. When the temperature is around 1 K, dark counts is no longer negligible. In this case, narrowing down the detection window can help to suppress the dark counts, however it is limited by the required time to keep the high fidelity of state transfer. It is noted that our proposed system exhibits great performance with extremely low dark counts and near unit fidelity under 100 mK. At the typical working temperature (50 mK), the dark count from thermal excitation is below the level 0.001, which is negligible and thus allows a more relaxed limitation of readout time window.

\begin{figure}[t]
	\centering
	\includegraphics[width=0.48\textwidth]{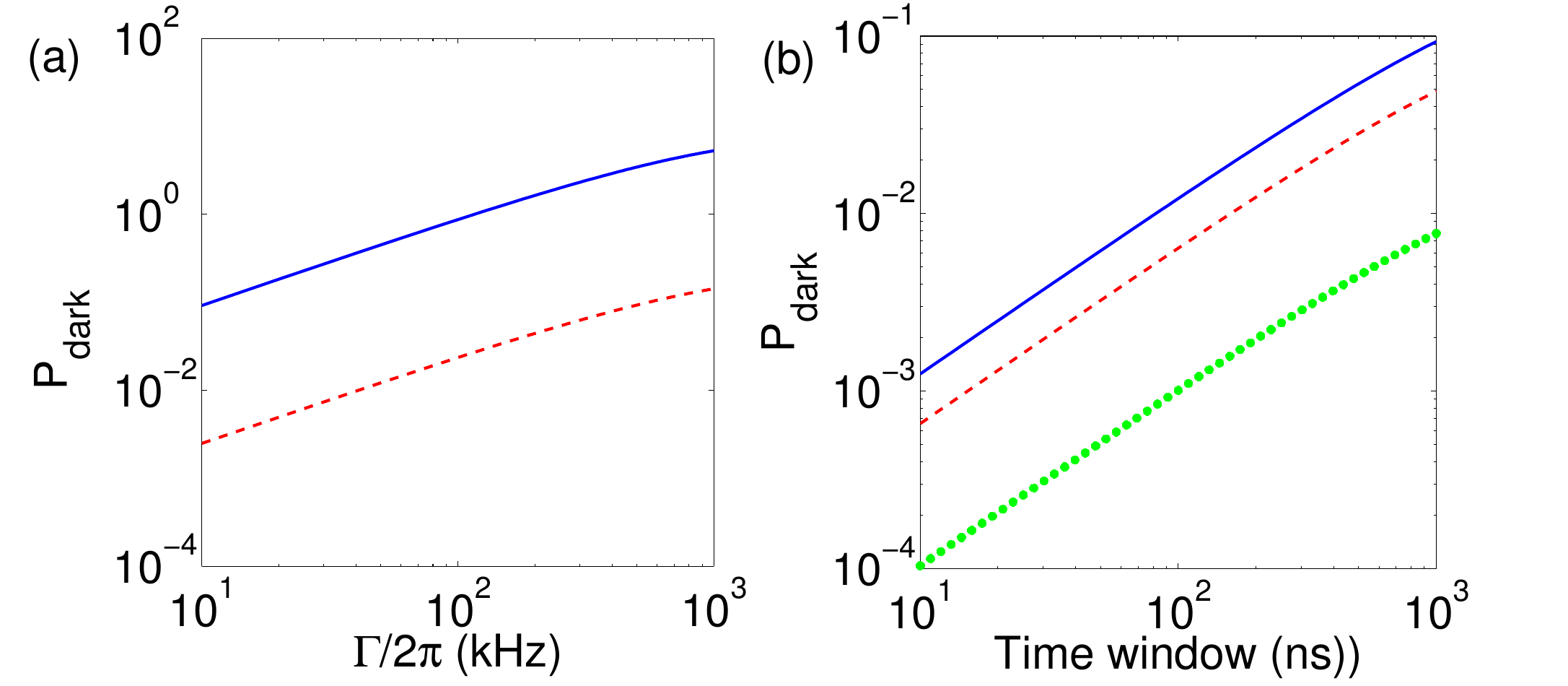}
	\caption{(Color online) (a) Dark count due to thermal excitation as a function of damping. Mechanical damping $\Gamma_{m}$ and electric damping $\Gamma_{LC}$ are tuned together while the readout time window is fixed at 200 ns. Two different temperatures are studied: red dash line and blue solid line respectively correspond to 100 mK and 1 K. (a) Dark count as a function of readout time windows at low temperatures. Three different temperature are plotted: 100 mK (blue solid curve), 80 mK (red dash curve) and 50 mK (thick green dot curve). At 50 mK, the dark count ratio is lower than 0.01 within a large time scale which is already quite small. Other parameters are the same as in Fig.(\ref{fig:darkth})}\label{fig:darkdamp}
\end{figure}

\begin{figure}[h]
	\centering
	\begin{minipage}{0.5\textwidth}
		\centering
		\includegraphics[scale=0.3]{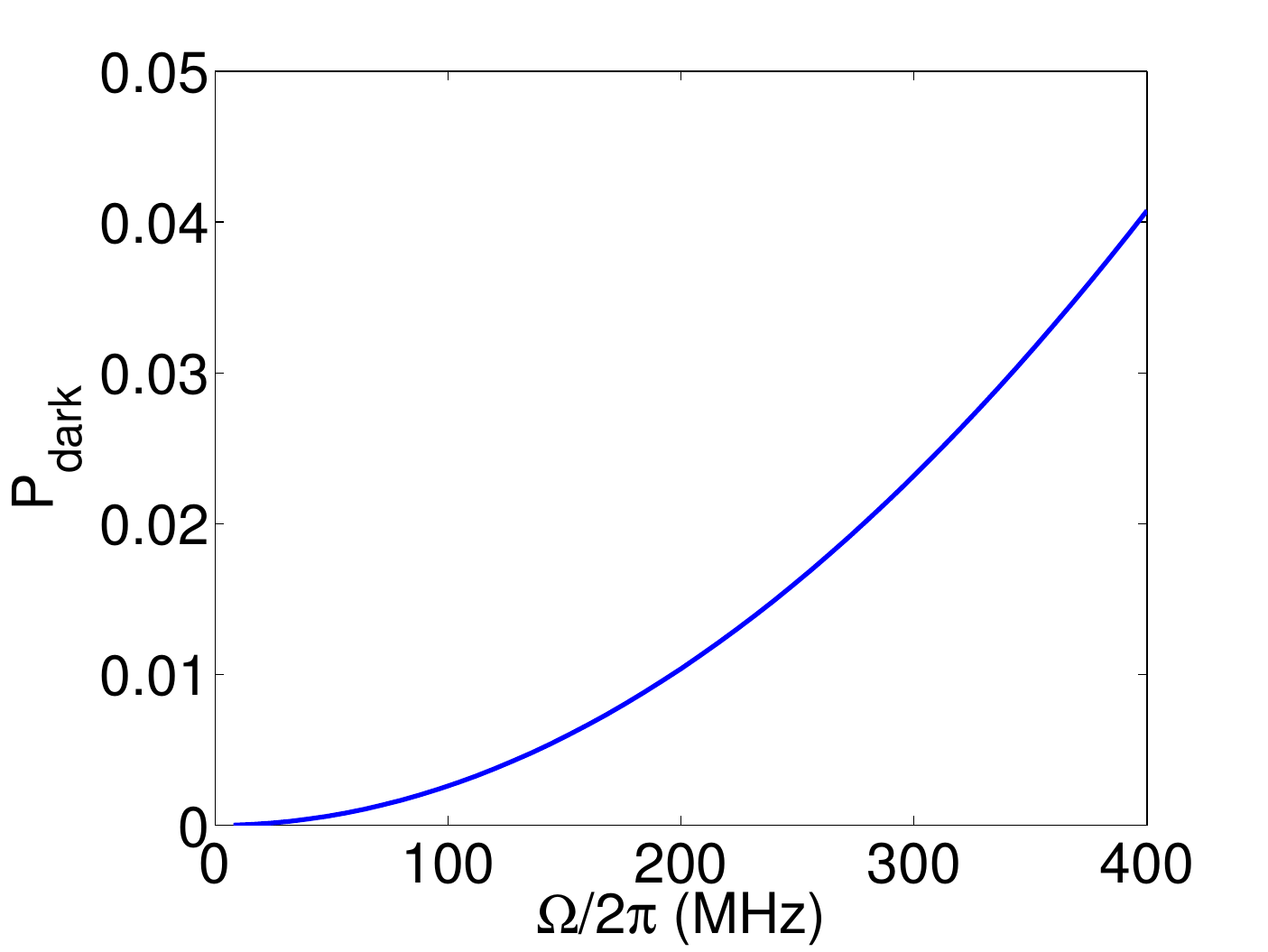}
	\end{minipage}

	\caption{(Color online) Here we present the dark count from the spontaneous emission from the excited state due to red-detuned laser driving. The time window is set at 200 ns as above. For $\Omega<190$ MHz, the dark count is lower than 0.01. Other parameters are taken as $\kappa/2\pi$=40 MHz, $\Gamma_{m}/2\pi$=10 kHz, $\Gamma_{LC}/2\pi$=10 kHz, $T=$ 50 mK, $\Delta/2\pi$=5 GHz.}\label{fig:darkspon}
\end{figure}

The second source of imperfection is the possibility of spontaneous emission from the excited state under red-detuned laser driving.  Its effects can be accounted by adding the non-resonant driving terming $\hbar\frac{\Omega}{2}(e^{-i\Delta t}$ $\ket{e}\bra{g}+H.c.)$ back to the Hamiltonian (\ref{dd}) and do the similar simulation as in Section \ref{sec:Fidelity} (see Fig. (\ref{fig:darkspon})). For simplicity, here we replaced decay terms $i\hbar\sqrt{\kappa/2\pi}\int_{-\omega_{a}}^{\omega_{a}	 }\\ 
d\omega(\ket{e}\bra{g}a(\omega)-\ket{g}\bra{e}a^\dagger(\omega))+\int_{-\omega_{a}}^{\omega_{a}}d\omega$ $[\hbar\omega a^\dagger(\omega)a(\omega)\\
-i\hbar\frac{\Gamma_{m}}{2}b^\dagger b-i\hbar\frac{\Gamma_{LC}}{2}c^\dagger c]$ by the decay term $-i\hbar\frac{\kappa}{2}\ket{e}\bra{e}$ (as we have $\kappa\gg \Gamma_{m},\Gamma_{LC}$). Then we define the dark count probability as $P_{dark}=1-P_n$, where $P_n$ was defined in Eq.(\ref{Pn}). Considering the time window $\tau$, the probability for a spontaneous emission can also be estimated by $\kappa\Omega^2/(8\Delta^2)\tau$ \cite{cirac1997quantum}. With a readout time window of 200 ns,  the dark count is lower than 0.01 when the Rabi frequency is smaller than 190 MHz. When the Rabi frequency is about 400 MHz, the dark count  is around 0.04. The dark counts given by this source is much larger than those caused by the former two sources. Thus the final fidelity is around 0.96 when the readout time window is 200 ns. To further reduce imperfections from the excitation of the driving laser, parameters of the sizes and bias voltage can be tuned to make the bare optomechanical coupling strength $g_{om}$ larger. Hence our proposed system as a quantum transducer maintains good performance with fidelity larger than 0.95 and transduction time less than 200 ns.

\section{\label{sec:level7}Conclusion}
In this paper, we propose a scheme to realize high speed quantum state transfer between a microwave photon in a LC circuit and an optical photon in free space with a single photon emitter in an atomic-thin 2D material. The 2D mechanical resonator couples an electric microwave photon to atomic ZPL with a coupling strength larger than 100~MHz. The conversion to a single photon pulse output can be realized by applying a weak driving laser at the red sideband of the atomic transition of the single-photon emitter. High speed  quantum state transfer between a single microwave photon and a single optical photon with a fidelity larger than 0.95 can be realized at 50 mK.

\begin{appendices}
	\appendix
	
	\section{\label{sec:level9}Hamiltonian of electro-optomechanical coupling}
	
	The Hamiltonian of the proposed electro-optomechanical system can be writen as \cite{14,13,16}
	\begin{align}
	\begin{split}
	H=&\hbar \omega_{0}\ket{e}\bra{e}+\frac{m\Omega^2x^2}{2}+\frac{p ^2}{2m}\\
	&+\frac{q ^2}{2C(x)}+ \frac{\phi ^2}{2L}-qV_{dc}.
	\end{split}
	\end{align}
	
	The corresponding Langevin equations are
	\begin{subequations}
		\begin{align}
		\dot{x}=&\frac{p}{m},\\
		\begin{split}
		\dot{p}=&-m\omega_{m}^{2}x-\frac{q^{2}}{2}\frac{\partial }{\partial x}(\frac{1}{C(x)})-\hbar\frac{\partial \omega_{0}(x)}{\partial x}\ket{e}\bra{e}\\
		&-\Gamma_{m}p-F_{th},
		\end{split}\\
		\dot{q}=&\frac{\phi}{L},\\
		\dot{\phi}=&-\frac{q}{C(x)}-\Gamma_{LC}\phi+V_{dc}+V_{th}.
		\end{align}
	\end{subequations}
	where $\Gamma_m$ and $\Gamma_{LC}$ refer to the dissipation rates of mechanical membrane and the LC circuit. Here $F_{th}$ is the thermal noise force and $V_{th}$ is the thermal noise voltage. The equilibrium state is then characterized by
	\begin{subequations}
		\begin{align}
		\bar{\phi}&=\bar{p}=0,\\
		\begin{split}
		m\omega_{m}^2\bar{x}&=-\frac{\bar{q}^{2}}{2}\frac{\partial }{\partial x}(\frac{1}{C(x)})|_{x=\bar{x}}-\hbar\frac{\partial \omega_{0}(\bar{x})}{\partial x}\ket{e}\bra{e}\\
		&\simeq -\frac{\bar{q}^{2}}{2}\frac{\partial }{\partial x}(\frac{1}{C(x)})|_{x=\bar{x}},
		\end{split}\\
		\bar{q}&=V_{dc}C(\bar{x}).
		\end{align}
	\end{subequations}
	
	We rewrite the Hamiltonian using $x=\bar{x}+\delta x$, $q=\bar{q}+\delta q$, and obtain
	\begin{align}
	H&=\hbar[\omega{0}(\bar{x})+\frac{\partial\omega_{0}}{\partial x}|_{x=\bar{x}}\delta x]\ket{e}\bra{e} +\frac{\phi^2}{2L}+\frac{p^2}{2m} \notag\\
	&+\frac{m\omega_{m}^2(\bar{x}^2+2\bar{x}\delta x+\delta x^2)}{2} \notag\\
	&+\frac{(\bar{q}^2+2\bar{q}\delta q+\delta q^2)}{2C(x)}-(\bar{q}+\delta q)V_{dc} \notag\\
	&=\underset{H_{0}}{\underbrace{\hbar\omega_{0}(\bar x)\ket{e}\bra{e} +\frac{\phi^2}{2L}+\frac{p^2}{2m}+\frac{m\omega_{m}^2}{2}\delta x^2+\frac{\delta q^2}{2C(\bar{x})}}}\notag \\
	&+\underset{H_{om}}{\underbrace{\hbar\frac{\partial\omega_{0}}{\partial x}\ket{e}\bra{e}\delta x }}
	+\underset{H_{em}}{\underbrace{\bar{q}\frac{\partial }{\partial x}(\frac{1}{C(x))})|_{x=\bar{x}}\delta q\delta x}}\notag \\ &+\underset{H_{const}}{\underbrace{\frac{\bar{q}^2}{2C(\bar{x})}+\frac{m\omega_{m}^2\bar{x}^2}{2}-\bar{q}V_{dc}}}.
	\end{align}
	
	Then we replace $\delta x$, $p$, $\delta q$  and $\phi$ with the photon and phonon operators,
	\begin{align*}
	&\delta x=x_{zpf}(b+b^{\dagger}),\;p=im_{eff}\omega_{m}x_{zpf}(b-b^\dagger),\\
	&\delta q=q_{zpf}(c+c^{\dagger}),\;\phi=iL\omega_{LC}q_{zpf	}(c-c^\dagger).
	\end{align*}
	Eventually, by ignoring the constant term $H_{const}$, the Hamiltonian takes the form $H=H_{0}+H_{I}$, where
	\begin{subequations}
		\begin{align}
		H_{0}&=\hbar\omega_{0}\ket{e}\bra{e}+\hbar\omega_{m}b^\dagger b+\hbar\omega_{LC}c^\dagger c,\\
		H_{I}&=\hbar g_{om}(b+b^\dagger)\ket{e}\bra{e} +\hbar g_{em}(b+b^\dagger)(c+c^\dagger).
		\end{align}
	\end{subequations}
	Here $g_{om}=x_{zpf}(\partial \omega_{0}/\partial x)$ and $g_{em}=\bar{q}\frac{\partial }{\partial x}(\frac{1}{C(x)})x_{zpf}q_{zpf}$ are the optomechanical and electromechanical coupling strength, respectively.
	
	\section{\label{sec:level11}Dynamics under off-resonant laser driving}
	In the main text, to achieve state transfer from a electronic photon to excited state of a single photon emitter, we apply a red-detuned laser driving which is described by a Hamiltonian
	\begin{equation}
	H_{d}=\hbar\frac{\Omega}{2} (e^{-i\omega_{L}t}\ket{e}\bra{g}+e^{i\omega_{L}t}\ket{g}\bra{e}),
	\end{equation}
	where $\omega_{L}$ is the frequency of the driving laser and $\Omega$ is the Rabi frequency. Applying the Schrieffer-Wolff transformation
	\cite{17,18}
	\begin{equation}
	U=\exp[\frac{g_{om}}{\omega_{m}}(b^\dagger-b)\ket{e}\bra{e}]
	\end{equation}
	to the total Hamiltonian yields
	\begin{equation}
	\begin{split}
	\tilde{H}&=\hbar \omega_{0} \ket{e}\bra{e}+\hbar\omega_{m}b^\dagger b+\hbar\omega_{LC}c^\dagger c-\hbar\frac{g_{om}^2}{\omega_{m}}\ket{e}\bra{e}\\
	&-2\hbar\frac{ g_{em}g_{om}}{\omega_{m}}(c+c^\dagger)\ket{e}\bra{e}\\
	&+ \hbar\frac{\Omega}{2}(e^{-i\omega_{L}t}e^{\frac{g_{om}}{\omega_{m}}(b^\dagger-b)}\ket{e}\bra{g}+H.c.). \label{hamiltonian}
	\end{split}	
	\end{equation}

	Then transformed to the rotating frame with the frequency of the laser drive, the Hamiltonian $H=H_{0}+H_{I}$ is given by
	\begin{align}
	H_{0}&=-\hbar \Delta \ket{e}\bra{e}+\hbar\omega_{m}b^\dagger b+\hbar\omega_{LC}c^\dagger c, \\
	\begin{split}
	H_{I}&=\hbar g_{em}(b+b^\dagger)(c+c^\dagger)-2\hbar\frac{g_{om}g_{em}}{\omega_{m}}(c^\dagger+c)\ket{e}\bra{e}\\
	&+\hbar\frac{\Omega}{2}(e^{\frac{g_{om}}{\omega_{m}}(b^\dagger-b)}\ket{e}\bra{g}+H.c.)\label{ehI}.
	\end{split}	
	\end{align}
	Here $\Delta=\omega_{L}-\omega_{0}+\frac{g_{om}^2}{\omega_{m}}$ is the detuning between the driving laser and the optical resonance frequency.
	In the interaction picture, we rewrite the Hamiltonian by using rotating wave approximation (RWA) and obtain the effective form
	\begin{equation}
	\begin{split}
	\tilde{H_{I}}&=\hbar g_{em}(bc^\dagger+b^\dagger c)\\
	&-2\hbar\frac{g_{om}g_{em}}{\omega_{m}}(c^\dagger e^{-i\omega_{LC}}+ce^{i\omega_{LC}})\ket{e}\bra{e}\\
	&+\hbar\frac{\Omega}{2}(e^{-i\Delta t}e^{\frac{g_{om}}{\omega_{m}}(b^\dagger e^{i\omega_{m}t}-be^{-i\omega_{m}t})}\ket{e}\bra{g}+H.c.).
	\end{split}\label{eh2}
	\end{equation}
	
	We assume that the laser driving field is tuned to near the red phonon sideband of the $\bar{g}$ to $\bar{e}$ transition ($\Delta\approx -\omega_{m}=-\omega_{LC}$).  The last term in Eq.(\ref{eh2}) oscillates rapidly in the interaction picture. So its average influence on the energy level of atom excited state $\ket{e}$ can be ignored. Then, expanding $\tilde{H}_{I}$ in Eq.(\ref{eh2}) by $g/\omega_{m}$ and keeping only the near resonant terms, we can approximate the interaction Hamiltonian as
	\begin{equation}
	\begin{split}
	\tilde{H_{I}}=&\hbar g_{em}(bc^\dagger+b^\dagger c)+\hbar\frac{\Omega}{2}\frac{g_{om}}{\omega_{m}}(b e^{-i(\Delta+\omega_{m})t}\ket{e}\bra{g}+H.c.)\\
	=&\hbar \tilde{g}_{om}(b \sigma_{eg}+b^\dagger \sigma_{ge})+\hbar g_{em}(bc^\dagger+b^\dagger c),
	\end{split}	
	\end{equation}
	where $\tilde{g}_{om}=\frac{\Omega}{2}\frac{g_{om}}{\omega_{m}}$ is the effective coupling strength between the electronic state of the single photon emitter and the mechanical vibration mode. 
	
\end{appendices}

\section*{Acknowledgments}

T. Li acknowledges supports from the Purdue Quantum Science and Engineering Institute (PQSEI) seed grant, and the Purdue EFC research grant. Z.-Q. Yin is supported by National Natural Science Foundation of China under Grant No. 61771278 and Beijing Institute of Technology Research Fund Program for Young Scholars. 

\section*{Conflict of Interest}

The authors declare no conflict of interest.


\bibliographystyle{andp2012}
\bibliography{reference}

\providecommand{\WileyBibTextsc}{}
\let\textsc\WileyBibTextsc
\providecommand{\othercit}{}
\providecommand{\jr}[1]{#1}
\providecommand{\etal}{~et~al.}


\begin{thebibliography}{[10]}

\bibitem{2}
 \textsc{M.~Aspelmeyer},  \textsc{T.\,J. Kippenberg},  and
  \textsc{F.~Marquardt} \jr{Rev. Mod. Phys.} \textbf{86}(4), 1391 (2014).


\bibitem{3}
 \textsc{A.\,H. Safavi-Naeini} and  \textsc{O.~Painter} \jr{New Journal of
  Physics} \textbf{13}(1), 013017 (2011).


\bibitem{4}
 \textsc{K.~Stannigel},  \textsc{P.~Rabl},  \textsc{A.\,S. S{\o}rensen},
  \textsc{P.~Zoller},  and  \textsc{M.\,D. Lukin} \jr{Physical review letters}
  \textbf{105}(22), 220501 (2010).


\bibitem{Bochmann2013}
 \textsc{J.~Bochmann},  \textsc{A.~Vainsencher},  \textsc{D.\,D. Awschalom},
  and  \textsc{A.\,N. Cleland} \jr{Nature Physics} \textbf{9}, 712--716 (2013).


\bibitem{32}
 \textsc{R.\,W. Andrews},  \textsc{R.\,W. Peterson},  \textsc{T.\,P. Purdy},
  \textsc{K.~Cicak},  \textsc{R.\,W. Simmonds},  \textsc{C.\,A. Regal},  and
  \textsc{K.\,W. Lehnert} \jr{Nature Physics} \textbf{10}(4), 321--326 (2014).


\bibitem{33}
 \textsc{T.~Bagci},  \textsc{A.~Simonsen},  \textsc{S.~Schmid},  \textsc{L.\,G.
  Villanueva},  \textsc{E.~Zeuthen},  \textsc{J.~Appel},  \textsc{J.\,M.
  Taylor},  \textsc{A.~S{\o}rensen},  \textsc{K.~Usami},
  \textsc{A.~Schliesser},  and  \textsc{E.\,S. Polzik} \jr{Nature}
  \textbf{507}(7490), 81--85 (2014).


\bibitem{34}
 \textsc{K.\,C. Balram},  \textsc{M.\,I. Davan{\c{c}}o},  \textsc{J.\,D. Song},
   and  \textsc{K.~Srinivasan} \jr{Nature photonics} \textbf{10}(5), 346--352
  (2016).


\bibitem{35}
 \textsc{L.~Tian} \jr{Annalen der Physik} \textbf{527}(1-2), 1--14 (2015).


\bibitem{36}
 \textsc{Z.\,q. Yin},  \textsc{W.~Yang},  \textsc{L.~Sun},  and
  \textsc{L.~Duan} \jr{Physical Review A} \textbf{91}(1), 012333 (2015).


\bibitem{yin2015b}
 \textsc{Z.\,Q. Yin},  \textsc{N.~Zhao},  and  \textsc{T.~Li} \jr{Sci
  China-Phys Mech Astron} \textbf{58}(5), 1--12 (2015).


\bibitem{13}
 \textsc{T.~Bagci},  \textsc{A.~Simonsen},  \textsc{S.~Schmid},  \textsc{L.\,G.
  Villanueva},  \textsc{E.~Zeuthen},  \textsc{J.~Appel},  \textsc{J.\,M.
  Taylor},  \textsc{A.~S{\o}rensen},  \textsc{K.~Usami},
  \textsc{A.~Schliesser},  and  \textsc{E.\,S. Polzik} \jr{Nature}
  \textbf{507}(7490), 81--85 (2014).


\bibitem{16}
 \textsc{J.\,M. Taylor},  \textsc{A.\,S. S{\o}rensen},  \textsc{C.\,M. Marcus},
   and  \textsc{E.\,S. Polzik} \jr{Physical Review Letters} \textbf{107}(27),
  273601 (2011).


\bibitem{27}
 \textsc{Y.\,M. He},  \textsc{G.~Clark},  \textsc{J.\,R. Schaibley},
  \textsc{Y.~He},  \textsc{M.\,C. Chen},  \textsc{Y.\,J. Wei},
  \textsc{X.~Ding},  \textsc{Q.~Zhang},  \textsc{W.~Yao},  \textsc{X.~Xu},
  \textsc{C.\,Y. Lu},  and  \textsc{J.\,W. Pan} \jr{Nature nanotechnology}
  \textbf{10}(6), 497--502 (2015).


\bibitem{Vamivakas2015}
 \textsc{C.~Chakraborty},  \textsc{L.~Kinnischtzke},  \textsc{K.~Goodfellow},
  \textsc{R.~Beams},  and  \textsc{A.\,N. A.~N.~Vamivakas} \jr{Nature
  nanotechnology} \textbf{10}(6), 507–511 (2015).


\bibitem{Imamoglu2015}
 \textsc{A.~Srivastava},  \textsc{M.~Sidler},  \textsc{A.\,V. Allain},
  \textsc{D.\,S. Lembke},  \textsc{A.~Kis},  and  \textsc{A.~Imamo{\v{g}}lu}
  \jr{Nature nanotechnology} \textbf{10}(6), 491 -- 496 (2015).


\bibitem{Potemski2015}
 \textsc{M.~Koperski},  \textsc{K.~Nogajewski},  \textsc{A.~Arora},
  \textsc{V.~Cherkez},  \textsc{P.~Mallet},  \textsc{J.\,Y. Veuillen},
  \textsc{J.~Marcus},  \textsc{P.~Kossacki},  and  \textsc{M.~Potemski}
  \jr{Nature nanotechnology} \textbf{10}(6), 503 -- 506 (2015).


\bibitem{26}
 \textsc{T.\,T. Tran},  \textsc{K.~Bray},  \textsc{M.\,J. Ford},
  \textsc{M.~Toth},  and  \textsc{I.~Aharonovich} \jr{Nature nanotechnology}
  \textbf{11}(1), 37--41 (2016).


\bibitem{7}
 \textsc{G.~Grosso},  \textsc{H.~Moon},  \textsc{B.~Lienhard},
  \textsc{S.~Ali},  \textsc{D.\,K. Efetov},  \textsc{M.\,M. Furchi},
  \textsc{P.~Jarillo-Herrero},  \textsc{M.\,J. Ford},  \textsc{I.~Aharonovich},
   and  \textsc{D.~Englund} \jr{Nat. Comm.} \textbf{8}, 705 (2017).


\bibitem{14}
 \textsc{K.\,W. Lee},  \textsc{D.~Lee},  \textsc{P.~Ovartchaiyapong},
  \textsc{J.~Minguzzi},  \textsc{J.\,R. Maze},  and  \textsc{A.\,C.\,B. Jayich}
  \jr{Physical Review Applied} \textbf{6}(3), 034005 (2016).


\bibitem{17}
 \textsc{D.\,A. Golter},  \textsc{T.~Oo},  \textsc{M.~Amezcua},  \textsc{K.\,A.
  Stewart},  and  \textsc{H.~Wang} \jr{Physical review letters}
  \textbf{116}(14), 143602 (2016).


\bibitem{18}
 \textsc{D.\,A. Golter},  \textsc{T.~Oo},  \textsc{M.~Amezcua},
  \textsc{I.~Lekavicius},  \textsc{K.\,A. Stewart},  and  \textsc{H.~Wang}
  \jr{Physical Review X} \textbf{6}(4), 041060 (2016).


\bibitem{57}
 \textsc{Y.~Zhou},  \textsc{G.~Scuri},  \textsc{J.~Sung},  \textsc{R.\,J.
  Gelly},  \textsc{D.\,S. Wild},  \textsc{K.~De~Greve},  \textsc{A.\,Y. Joe},
  \textsc{T.~Taniguchi},  \textsc{K.~Watanabe},  \textsc{P.~Kim},
  \textsc{M.\,D. Lukin},  and  \textsc{H.~Park} \jr{Phys. Rev. Lett.}
  \textbf{124}(Jan), 027401 (2020).


\bibitem{Rabl2009}
 \textsc{P.~Rabl},  \textsc{P.~Cappellaro},  \textsc{M.\,V.\,G. Dutt},
  \textsc{L.~Jiang},  \textsc{J.\,R. Maze},  and  \textsc{M.\,D. Lukin}
  \jr{Phys. Rev. B} \textbf{79}(Jan), 041302 (2009).


\bibitem{31}
 \textsc{M.~Abdi},  \textsc{M.\,J. Hwang},  \textsc{M.~Aghtar},  and
  \textsc{M.\,B. Plenio} \jr{Physical Review Letters} \textbf{119}(23), 233602
  (2017).


\bibitem{Yin2013}
 \textsc{Z.\,q. Yin},  \textsc{T.~Li},  \textsc{X.~Zhang},  and  \textsc{L.\,M.
  Duan} \jr{Phys. Rev. A} \textbf{88}(Sep), 033614 (2013).


\bibitem{Ma2016}
 \textsc{Y.~Ma},  \textsc{Z.\,q. Yin},  \textsc{P.~Huang},  \textsc{W.\,L.
  Yang},  and  \textsc{J.~Du} \jr{Phys. Rev. A} \textbf{94}(Nov), 053836
  (2016).


\bibitem{Ma2017}
 \textsc{Y.~Ma},  \textsc{T.\,M. Hoang},  \textsc{M.~Gong},  \textsc{T.~Li},
  and  \textsc{Z.\,q. Yin} \jr{Phys. Rev. A} \textbf{96}(Aug), 023827 (2017).


\bibitem{li2020preparing}
 \textsc{B.~Li},  \textsc{X.~Li},  \textsc{P.~Li},  and  \textsc{T.~Li}
  \jr{Advanced Quantum Technologies} \textbf{3}(6), 2000034 (2020).


\bibitem{xu2019quantum}
 \textsc{Z.~Xu},  \textsc{Z.\,q. Yin},  \textsc{Q.~Han},  and  \textsc{T.~Li}
  \jr{Optical Materials Express} \textbf{9}(12), 4654--4668 (2019).


\bibitem{chen2019nonadiabatic}
 \textsc{X.\,Y. Chen},  \textsc{T.~Li},  and  \textsc{Z.\,Q. Yin} \jr{Science
  Bulletin} \textbf{64}(6), 380--384 (2019).


\bibitem{kelly2015}
 \textsc{J.~Kelly} \etal{} \jr{Nature} \textbf{519}, 66--69 (2015).


\bibitem{zheng2017}
 \textsc{Y.~Zheng} \etal{} \jr{Physical review letters} \textbf{118}, 210504
  (2017).


\bibitem{sontheimer2017photodynamics}
 \textsc{B.~Sontheimer},  \textsc{M.~Braun},  \textsc{N.~Nikolay},
  \textsc{N.~Sadzak},  \textsc{I.~Aharonovich},  and  \textsc{O.~Benson}
  \jr{Physical Review B} \textbf{96}(12), 121202 (2017).


\bibitem{dietrich2017narrowband}
 \textsc{A.~Dietrich},  \textsc{M.~B\"urk},  \textsc{E.\,S. Steiger},
  \textsc{L.~Antoniuk},  \textsc{T.\,T. Tran},  \textsc{M.~Nguyen},
  \textsc{I.~Aharonovich},  \textsc{F.~Jelezko},  and  \textsc{A.~Kubanek}
  \jr{Phys. Rev. B} \textbf{98}(Aug), 081414 (2018).


\bibitem{19}
 \textsc{D.~Lee},  \textsc{K.\,W. Lee},  \textsc{J.\,V. Cady},
  \textsc{P.~Ovartchaiyapong},  and  \textsc{A.\,C.\,B. Jayich} \jr{Journal of
  Optics} \textbf{19}(3), 033001 (2017).


\bibitem{5}
 \textsc{A.~Castellanos-Gomez},  \textsc{V.~Singh},  \textsc{H.\,S. van\,der
  Zant},  and  \textsc{G.\,A. Steele} \jr{Annalen der Physik}
  \textbf{527}(1-2), 27--44 (2015).


\bibitem{will2017high}
 \textsc{M.~Will},  \textsc{M.~Hamer},  \textsc{M.~Muller},  \textsc{A.~Noury},
   \textsc{P.~Weber},  \textsc{A.~Bachtold},  \textsc{R.~Gorbachev},
  \textsc{C.~Stampfer},  and  \textsc{J.~Guttinger} \jr{Nano letters}
  \textbf{17}(10), 5950--5955 (2017).


\bibitem{lee2019ultrahigh}
 \textsc{M.~Lee} \jr{Japanese Journal of Applied Physics} \textbf{58}(10),
  100914 (2019).


\bibitem{11}
 \textsc{A.~Eichler},  \textsc{J.~Moser},  \textsc{J.~Chaste},
  \textsc{M.~Zdrojek},  \textsc{I.~Wilson-Rae},  and  \textsc{A.~Bachtold}
  \jr{Nature nanotechnology} \textbf{6}(6), 339--342 (2011).


\bibitem{12}
 \textsc{X.~Song},  \textsc{M.~Oksanen},  \textsc{J.~Li},  \textsc{P.~Hakonen},
   and  \textsc{M.\,A. Sillanp{\"a}{\"a}} \jr{Physical review letters}
  \textbf{113}(2), 027404 (2014).


\bibitem{60}
 \textsc{D.~Wang},  \textsc{H.~Kelkar},  \textsc{D.~Martin-Cano},
  \textsc{D.~Rattenbacher},  \textsc{A.~Shkarin},  \textsc{T.~Utikal},
  \textsc{S.~G{\"{o}}tzinger},  and  \textsc{V.~Sandoghdar} \jr{Nature Physics}
  \textbf{15}(5), 483--489 (2019).


\bibitem{61}
 \textsc{K.~Lee},  \textsc{X.~Chen},  \textsc{H.~Eghlidi},  \textsc{P.~Kukura},
   \textsc{R.~Lettow},  \textsc{A.~Renn},  \textsc{V.~Sandoghdar},  and
  \textsc{S.~G{\"o}tzinger} \jr{Nature Photonics} \textbf{5}(3), 166 (2011).


\bibitem{hunger2010fiber}
 \textsc{D.~Hunger},  \textsc{T.~Steinmetz},  \textsc{Y.~Colombe},
  \textsc{C.~Deutsch},  \textsc{T.\,W. H{\"a}nsch},  and  \textsc{J.~Reichel}
  \jr{New Journal of Physics} \textbf{12}(6), 065038 (2010).


\bibitem{snijders2018fiber}
 \textsc{H.~Snijders},  \textsc{J.~Frey},  \textsc{J.~Norman},
  \textsc{V.~Post},  \textsc{A.~Gossard},  \textsc{J.~Bowers},  \textsc{M.~van
  Exter},  \textsc{W.~L{\"o}ffler},  and  \textsc{D.~Bouwmeester} \jr{Physical
  Review Applied} \textbf{9}(3), 031002 (2018).


\bibitem{ahn2018stable}
 \textsc{J.~Ahn},  \textsc{Z.~Xu},  \textsc{J.~Bang},  \textsc{A.\,E.\,L.
  Allcca},  \textsc{Y.\,P. Chen},  and  \textsc{T.~Li} \jr{Optics letters}
  \textbf{43}(15), 3778--3781 (2018).


\bibitem{6}
 \textsc{C.~Lee},  \textsc{X.~Wei},  \textsc{J.\,W. Kysar},  and
  \textsc{J.~Hone} \jr{Science} \textbf{321}(5887), 385--388 (2008).


\bibitem{15}
 \textsc{C.~Wong},  \textsc{M.~Annamalai},  \textsc{Z.~Wang},  and
  \textsc{M.~Palaniapan} \jr{Journal of Micromechanics and Microengineering}
  \textbf{20}(11), 115029 (2010).


\bibitem{38}
 \textsc{N.~Morell},  \textsc{A.~Reserbat-Plantey},  \textsc{I.~Tsioutsios},
  \textsc{K.\,G. Schädler},  \textsc{F.~Dubin},  \textsc{F.\,H. Koppens},
  and  \textsc{A.~Bachtold} \jr{Nano letters} \textbf{16}(8), 5102--5108
  (2016).


\bibitem{39}
 \textsc{P.~Weber},  \textsc{J.~Guttinger},  \textsc{I.~Tsioutsios},
  \textsc{D.\,E. Chang},  and  \textsc{A.~Bachtold} \jr{Nano letters}
  \textbf{14}(5), 2854--2860 (2014).


\bibitem{8}
 \textsc{C.~Chakraborty},  \textsc{K.\,M. Goodfellow},  \textsc{S.~Dhara},
  \textsc{A.~Yoshimura},  \textsc{V.~Meunier},  and  \textsc{A.\,N. Vamivakas}
  \jr{Nano Letters} \textbf{17}(4), 2253--2258 (2017).


\bibitem{xia2019room}
 \textsc{Y.~Xia},  \textsc{Q.~Li},  \textsc{J.~Kim},  \textsc{W.~Bao},
  \textsc{C.~Gong},  \textsc{S.~Yang},  \textsc{Y.~Wang},  and
  \textsc{X.~Zhang} \jr{Nano letters} \textbf{19}(10), 7100--7105 (2019).


\bibitem{ye2017}
 \textsc{F.~Ye},  \textsc{J.~Lee},  and  \textsc{P.\,X.\,L. Feng}
  \jr{Nanoscale} \textbf{9}, 18208 (2017).


\bibitem{25}
 \textsc{A.~Reserbat-Plantey},  \textsc{K.\,G. Sch{\"a}dler},
  \textsc{L.~Gaudreau},  \textsc{G.~Navickaite},  \textsc{J.~G{\"u}ttinger},
  \textsc{D.~Chang},  \textsc{C.~Toninelli},  \textsc{A.~Bachtold},  and
  \textsc{F.\,H. Koppens} \jr{Nature communications} \textbf{7}, 10218 (2016).


\bibitem{9}
 \textsc{J.\,S. Bunch},  \textsc{A.\,M. Van Der~Zande},  \textsc{S.\,S.
  Verbridge},  \textsc{I.\,W. Frank},  \textsc{D.\,M. Tanenbaum},
  \textsc{J.\,M. Parpia},  \textsc{H.\,G. Craighead},  and  \textsc{P.\,L.
  McEuen} \jr{Science} \textbf{315}(5811), 490--493 (2007).


\bibitem{23}
 \textsc{A.~Neto} and  \textsc{K.~Novoselov} \jr{Materials Express}
  \textbf{1}(1), 10--17 (2011).


\bibitem{24}
 \textsc{Q.\,H. Wang},  \textsc{K.~Kalantar-Zadeh},  \textsc{A.~Kis},
  \textsc{J.\,N. Coleman},  and  \textsc{M.\,S. Strano} \jr{Nature
  nanotechnology} \textbf{7}(11), 699--712 (2012).


\bibitem{falin2017mechanical}
 \textsc{A.~Falin},  \textsc{Q.~Cai},  \textsc{E.\,J. Santos},
  \textsc{D.~Scullion},  \textsc{D.~Qian},  \textsc{R.~Zhang},
  \textsc{Z.~Yang},  \textsc{S.~Huang},  \textsc{K.~Watanabe},
  \textsc{T.~Taniguchi} \etal{} \jr{Nature communications} \textbf{8}(15815),
  1--9 (2017).


\bibitem{29}
 \textsc{I.~Frank},  \textsc{D.\,M. Tanenbaum},  \textsc{A.\,M. van\,der
  Zande},  and  \textsc{P.\,L. McEuen} \jr{Journal of Vacuum Science \&
  Technology B: Microelectronics and Nanometer Structures Processing,
  Measurement, and Phenomena} \textbf{25}(6), 2558--2561 (2007).


\bibitem{30}
 \textsc{C.~Wong},  \textsc{M.~Annamalai},  \textsc{Z.~Wang},  and
  \textsc{M.~Palaniapan} \jr{Journal of Micromechanics and Microengineering}
  \textbf{20}(11), 115029 (2010).


\bibitem{37}
 \textsc{P.~Weber},  \textsc{J.~Guttinger},  \textsc{I.~Tsioutsios},
  \textsc{D.\,E. Chang},  and  \textsc{A.~Bachtold} \jr{Nano letters}
  \textbf{14}(5), 2854--2860 (2014).


\bibitem{40}
 \textsc{J.~Teufel},  \textsc{T.~Donner},  \textsc{D.~Li},  \textsc{J.~Harlow},
   \textsc{M.~Allman},  \textsc{K.~Cicak},  \textsc{A.~Sirois},  \textsc{J.\,D.
  Whittaker},  \textsc{K.~Lehnert},  and  \textsc{R.\,W. Simmonds} \jr{Nature}
  \textbf{475}(7356), 359--363 (2011).


\bibitem{10}
 \textsc{L.\,M. Duan},  \textsc{A.~Kuzmich},  and  \textsc{H.~Kimble}
  \jr{Physical Review A} \textbf{67}(3), 032305 (2003).


\bibitem{RevModPhys.70.101}
 \textsc{M.\,B. Plenio} and  \textsc{P.\,L. Knight} \jr{Rev. Mod. Phys.}
  \textbf{70}(Jan), 101--144 (1998).


\bibitem{PhysRevA.94.022302}
 \textsc{Y.~Huang},  \textsc{Z.\,q. Yin},  and  \textsc{W.\,L. Yang} \jr{Phys.
  Rev. A} \textbf{94}(Aug), 022302 (2016).


\bibitem{cirac1997quantum}
 \textsc{J.\,I. Cirac},  \textsc{P.~Zoller},  \textsc{H.\,J. Kimble},  and
  \textsc{H.~Mabuchi} \jr{Physical Review Letters} \textbf{78}(16), 3221
  (1997).


\end{thebibliography}

\end{document}